\documentclass[prb, reprint, 9pt, superscriptaddress,notitlepage, nofootinbib, longbibliography,
floatfix]{revtex4-2}

\usepackage{natbib}
\usepackage{graphicx}
\usepackage{epsfig}
\usepackage{amsfonts} 
\usepackage{mathtools}
\usepackage{amsmath} 
\usepackage{amssymb}
\usepackage{diagbox}
\usepackage{dsfont}
\usepackage[dvipsnames]{xcolor}
\usepackage{bm}
\usepackage{braket}
\usepackage{color}
\usepackage{physics} 
\usepackage{bbm}
\usepackage{hyperref}
\usepackage{subcaption}
\usepackage{comment}
\usepackage[normalem]{ulem}  
\usepackage[export]{adjustbox}  
\usepackage{enumitem}
\usepackage[titletoc, title]{appendix}
\usepackage{titletoc}
\usepackage{amsfonts}

\renewcommand{\raggedright}{\leftskip=0pt \rightskip=0pt plus 0cm}
\newcommand{\be}{\begin{equation}}
	\newcommand{\ee}{\end{equation}}
\newcommand{\ba}{\begin{eqnarray}}
	\newcommand{\ea}{\end{eqnarray}}

\definecolor{LinkColor}{rgb}{0,0,1}
\hypersetup{
	colorlinks=true,
	citecolor=LinkColor,
	linkcolor=LinkColor,
	urlcolor=LinkColor
}

\definecolor{gr}{rgb}{0,0,0}

\begin{document}

\title{Non-onsite symmetry breaking: topological phase coexistence and criticality}

		\author{Zhehao Zhang}\email{zhehao@umail.ucsb.edu}
			\affiliation{Department of Physics, University of California, Santa Barbara, CA 93106, USA}

		\author{Yabo Li}\email{liyb.poiuy@gmail.com}
			\affiliation{Center for Quantum Phenomena, Department of Physics, New York University, 726 Broadway, New York, New York 10003, USA}
		
		\author{Tsung-Cheng Lu}\email{tclu@umd.edu}
			\affiliation{Joint Center for Quantum Information and Computer Science, University of Maryland, College Park, Maryland 20742, USA}

\begin{abstract}
We explore the states of matter arising from the spontaneous symmetry breaking (SSB) of $\mathbb{Z}_2$ non-onsite symmetries. In one spatial dimension, we construct a frustration-free lattice model exhibiting SSB of a non-onsite symmetry, which features the coexistence of two ground states with distinct symmetry-protected topological (SPT) orders. We analytically prove the two-fold ground-state degeneracy and the existence of a finite energy gap. Fixing the symmetry sector yields a long-range entangled ground state that features long-range correlations among non-invertible charged operators. We also present a constant-depth measurement-feedback protocol to prepare such a state with a constant success probability in the thermodynamic limit, which may be of independent interest. Under a symmetric deformation, the SSB persists up to a critical point, beyond which a gapless phase characterized by a conformal field theory emerges. In two spatial dimensions, the SSB of 1-form non-onsite symmetries leads to a long-range entangled state (SPT soup) - a condensate of 1d SPT along any closed loops. On a torus, there are four such locally indistinguishable states that exhibit algebraic correlations between local operators, which we derived via a mapping to the critical $O(2)$ loop model. This constitutes an intriguing example of `topological quantum criticality'.

\end{abstract}

\maketitle
		
	{
		\hypersetup{linkcolor=black}
	}

\section{Introduction}

Symmetry provides a guiding principle to characterize quantum phases of matter. A simple example is the $\mathbb{Z}_2$ global symmetry, e.g. $\prod_i X_i$ in the $d$-space dimensional transverse-field Ising model $- \sum_{\expval{ij}} Z_iZ_j  - g \sum_i X_i$, which exhibits an ordered (disordered) phase due to the presence (absence) of spontaneous symmetry breaking. To date, there has been substantial progress in generalizing the conventional global symmetry in various ways to describe various exotic quantum orders (see Ref.\cite{McGreevy_review_symmetry_2023} for a review).

In this work, we show that even the global $\mathbb{Z}_2$ symmetry, arguably the simplest symmetry, when realized in an unconventional manner, can lead to novel quantum states of matter. Specifically, we consider the \textit{non-onsite} realization of a $\mathbb{Z}_2$ symmetry (non-onsite symmetry for short) such as $U=\prod_i U_{i,i+1}$, which is composed by commuting two-qubit unitary operations and $U^2=1$. This is in contrast to the conventional symmetry realization, which is \textit{onsite} - symmetry generator composed of products of single-site unitary operators such as $\prod_i X_i$. One can also take a further step to define a higher-form non-onsite symmetry by considering overlapping multi-qubit unitary gates acting on deformable sub-dimensional manifolds. The key question that we will be exploring is: \textit{when these non-onsite symmetries are spontaneously broken, what states of matter may emerge?} As we will show below, the spontaneous non-onsite symmetry breaking leads to several exotic features, e.g. including a gapped phase with the coexistence of trivial/non-trivial SPT (symmetry-protected topological) order \cite{spt_1d_2011, spt_2011} in 1d, as well as four locally indistinguishable states with power-law correlations in 2d.

\begin{figure}[t!]
    \centering
    \includegraphics[width=1\linewidth]{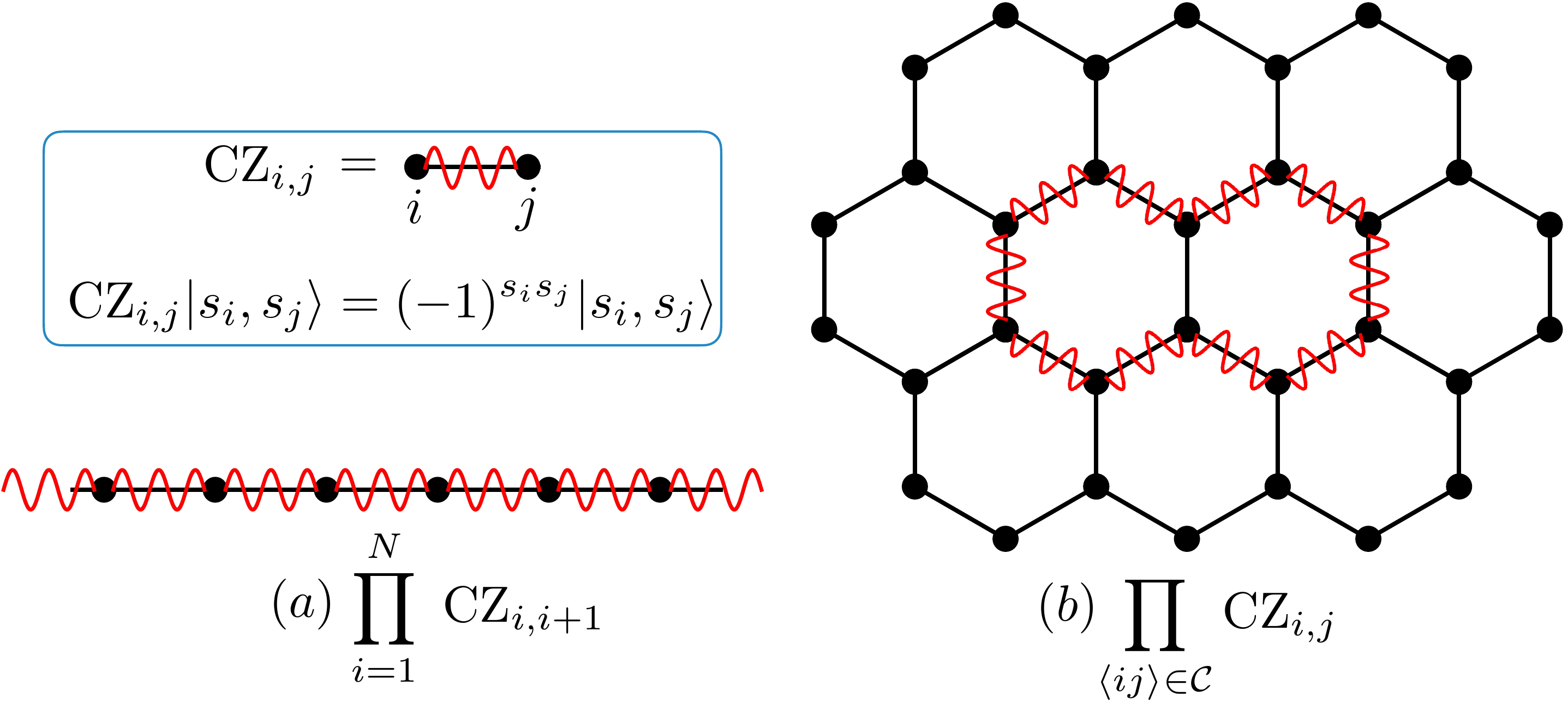}
    \caption{We discuss the spontaneous breaking of two types of non-onsite symmetry: (a) the global (0-form) symmetry consisting of controlled-Z gates in 1d and (b) the 1-form symmetry consisting of controlled-Z gates along any closed loops $\mathcal{C}$ in a 2d honeycomb lattice.} 
    \label{fig:symmetry}
\end{figure}

\section{Non-onsite symmetries}

We begin by properly defining a non-onsite symmetry. Given a lattice, we can equip it with a tensor-product Hilbert space by putting a qubit (or qudit) on each vertex (or link, plaquette, ...). A site is a grouping of neighboring qubits, and different sites do not overlap. Conventionally, symmetries are assumed to be onsite, i.e. the symmetry transformation of some symmetry group $\mathcal{G}$ can be implemented by a depth-1 circuit 
\begin{equation}
    U=\bigotimes_{\text{site }s}U_s.
\end{equation}
Each local gate $U_s$ is supported only on site $s$ \footnote{For instance, the unitary transformation $\prod_i U_{2i-1,2i}$ on a 1d lattice define an onsite symmetry since the two vertices $2i-1, 2i$ can be grouped to define a site.} and is a linear representation of $\mathcal{G}$. We call a symmetry non-onsite if its symmetry operator cannot be written as above under an arbitrary grouping of neighboring qubits. Note that an onsite unitary $U$ under the conjugation of a finite-depth local unitary circuit $V$ generically leads to a non-onsite symmetry $VUV^{\dagger}$, but the classifications of quantum phases under both symmetries are identical. We therefore focus on the \textit{intrinsically} non-onsite symmetries, which cannot be brought onsite by conjugation of any (finite-depth) unitary operator\footnote{We note that (i) performing a duality transformation or (ii) adding ancillae can bring an intrinsically non-onsite symmetry into an onsite symmetry. See Appendix.\ref{1d:remark} for details.}.

\section{Non-onsite symmetry breaking in 1d}
As a simple starting point, we consider a periodic 1d lattice consisting of $N$ qubits with 
even $N$, and define a global non-onsite $\mathbb{Z}_2$ symmetry $U_{\text{CZ}} = \prod_i CZ_{i,i+1}$, where  $CZ_{i,i+1}$ is the controlled-Z gate on two neighboring qubits on site $i$ and $i+1$. $U_{\text{CZ}}$ is intrinsically non-onsite since there does not exist a unitary operator that can transform $U_{\text{CZ}}$ to a product of non-overlapping local unitary gates (see Appendix.\ref{sec:prove_non_onsite} for proof). The key idea of the proof is to compute the trace $\tr U_{\text{CZ}}$, and show that it possesses a constant prefactor that distinguishes it from the the trace of any global $\mathbb{Z}_2$ onsite symmetries.

To explore the physics of spontaneous non-onsite symmetry breaking, we would like to construct two states that can be transformed into each other by $U_{\text{CZ}}$ and are both the ground states of a local Hamiltonian symmetric under $U_{\text{CZ}}$. One choice we can adopt is $\ket{+}^{\otimes N}$ and $\ket{\text{cluster}}= U_{\text{CZ}}\ket{+}^{\otimes N}$. The former is an X-basis product state and the latter is the so-called cluster state \cite{Raussendorf_2001_ghz}. They correspond to a trivial and a non-trivial SPT respectively under the $\mathbb{Z}_2\times  \mathbb{Z}_2$ symmetry generated by $\prod_{i\in o}X_i$ and $\prod_{i\in e}X_i$, with  $o/e$ denotes the set of odd/even sites \cite{spt_1d_2011, spt_2011}. The spontaneous non-onsite symmetry breaking therefore suggests the coexistence of trivial/non-trivial SPT orders in the ground subspace of a local Hamiltonian.

One question immediately arises: does there exist a gapped, local, $U_{\text{CZ}}$-symmetric Hamiltonian with the trivial/non-trivial SPT coexistence to manifest the spontaneous $U_{\text{CZ}}$ symmetric breaking? Notably, we construct such a Hamiltonian 
\footnote{Using the framework of MPS (matrix product state), Ref.\cite{stephen2024preparing} derived an alternative gapped parent non-onsite symmetric Hamiltonian with the two-fold degenerate ground states $\ket{+}^{\otimes N}$ and $\ket{\text{cluster}}$. The Hamiltonian reads $H= \sum_i h_i+ U_{\text{CZ}}h_iU_{\text{CZ}}  $ with $h_i=  (3-X_i )  (1-Z_{i-1}X_iZ_{i+1}) - (1+X_i) (X_{i-1}X_{i+1}+Y_{i-1}Y_{i+1})$. A major distinction between the two constructions is that our construction can be easily generalized to construct a parent Hamiltonian that exhibits the coexistence between any two Pauli stabilizer states in arbitrary spatial dimensions.} by drawing inspiration from the O’Brien-Fendley Hamiltonian \cite{o2018lattice}, which exhibits an order-disorder coexistence and spontaneous breaking of Kramer-Wannier duality symmetry \cite{shao_2024_LSM}. Our Hamiltonian reads

\begin{equation} \label{eq:1d_hamiltonian}
\begin{split}
    H &= \sum_{i} (1 - X_i) (1 - Z_{i+1}X_{i+2} Z_{i+3})\\
    &+ \sum_i (1 - Z_i X_{i+1} Z_{i+2}) (1 - X_{i+3})
\end{split}
\end{equation}
This Hamiltonian is frustration-free, and every local term annihilates both $\ket{+}^{\otimes N}$ and $\ket{\text{cluster}}$, so both of them are the ground states of $H$. Using the technique of Ref.\cite{duality_Huang_2024}, we analytically prove that these two states are the only ground states and there exists a finite energy gap in the thermodynamic limit (see Appendix.\ref{appendix: proof of GSD}, \ref{appendix:gap}). 

The two-fold degeneracy allows us to define a long-range entangled ground state $\ket{\psi} \propto \ket{+}^{\otimes N  }+ \ket{\text{cluster}}$ \footnote{This state has first appeared in Ref.\cite{stephen2024preparing,stephen2024many}, but the physics was not explored in details.}. $\ket{\psi}$ cannot be connected to a product state using finite-depth unitary circuits or constant-time adiabatic Hamiltonian evolution. To see this, one notices that there exists another state $\ket{\psi_-} \propto\ket{+}^{\otimes N  } - \ket{\text{cluster}}$, which is locally indistinguishable from $\ket{\psi}$ (see Appendix.\ref{sec:local indistinguishability} for proof). It follows that both of them cannot be connected to a product state via local unitary circuits, based on the Lieb-Robinson bound \cite{lieb1972finite,hastings_lrbound_2006}.


Is there an intrinsic physical property to manifest the long-range entanglement? For a long-range entangled state arising from spontaneous symmetry breaking, one essential feature is the long-range correlation among charged operators. A simple example is the GHZ state ($\propto \ket{0}^{\otimes N} + \ket{1}^{\otimes N}$) associated with the spontaneous breaking of the global $\mathbb{Z}_2$ symmetry $U_X=\prod_i X_i$, in which case the $\expval{Z_i Z_j} =1 $ for arbitrarily distant $i,j$. In particular, since $Z_i$ is charged under the symmetry $U_X$ (meaning  $U_X^{\dagger} Z_iU_X= -Z_i$), one has $\expval{Z_i}=0$. Consequently, $\expval{Z_i Z_j}$ is the same as the connected correlator $\expval{Z_i Z_j}-\expval{Z_i}\expval{Z_j}$, whose non-zero value directly diagnoses the long-range entanglement.

Similarly, for diagnosing the long-range order of the $\ket{\psi}$ arising from the spontaneous breaking of the non-onsite symmetry $U_{\text{CZ}}$, we can first find a corresponding local charged operator. Intriguingly, unlike the onsite symmetry $U_X$, which has a unitary charged operator, any valid charged operator of $U_{\text{CZ}}$ cannot be unitary (it is not even invertible) since the trace  $\tr U_{\text{CZ}} \neq 0$. To see this, imagine we have a charged operator $O$, meaning $U_{\text{CZ}} O U_{\text{CZ}}^{\dagger} = -O$. Assuming $O$ has an inverse, then one has $O^{-1} U_{\text{CZ}} O= -U_{\text{CZ}}$. Taking trace on both hand sides gives $\tr U_{CZ} = - \tr U_{CZ} =0$, leading to a contradiction since $\tr U_{CZ} \neq 0$. Therefore, $O$ does not have an inverse \footnote{Based on the trace of the symmetry generator, it is straightforward to show that a large class of SPT entanglers, when viewed as non-onsite symmetries, have non-invertible charged operators. However, having a non-invertible charge operator is not a unique feature of non-onsite symmetries. Certain onsite symmetries can also have a non-zero trace, thereby having a non-invertible charge operator. See Appendix.\ref{append:remark_charged_op} for details.}.  Indeed, we find a non-invertible charged operator $ O_i =X_i(1- Z_{i-1} Z_{i+1})= X_i -Z_{i-1} X_iZ_{i+1}$ with  $U_{\text{CZ}}^{\dagger}O_i U_{\text{CZ}} = -O_i$, which follows from $X_i \longleftrightarrow Z_{i-1} X_iZ_{i+1}$ under the conjugation of $U_{\text{CZ}}$. Via a straightforward calculation (Appendix.\ref{appendix:1d_order}), one finds $\expval{O_iO_j}_c= \expval{O_i O_j} - \expval{O_i } \expval{O_j} =1$ for any $i \neq j$ when $N\to \infty$, thereby diagnosing the long-range order. Such a long-range order also suggest a vanishing disorder parameter of $\ket{\psi}$, which we compute analytically in Appendix.\ref{appendix:1d_disorder}.

Does the spontaneous non-onsite symmetry breaking admit a stable gapped phase? Here we show the answer is affirmative by considering a certain non-onsite symmetric perturbation. Since the Hamiltonian $H$ has a $\mathbb{Z}_2^3$ symmetry generated by $\prod_{i\in o} X_i, \prod_{i\in e} X_i $ and the $U_{\text{CZ}}$, it is natural to consider a perturbation $V$ that preserves all these three symmetries: $V=\sum_i Z_{i}X_{i+1}Z_{i+3}X_{i+4} + \sum_i X_{i}Z_{i+1}X_{i+2}Z_{i+3}$. This leads to the deformed Hamiltonian $H' =  H + g V$ with $H = \sum_i (1 - X_i) (1 - Z_{i+1}X_{i+2} Z_{i+3})+ (1 - Z_i X_{i+1} Z_{i+2}) (1 - X_{i+3})$ defined in Eq.\ref{eq:1d_hamiltonian}. Alternatively, $H'$ may be expressed as (up to a constant)

\begin{equation}\label{eq:1d_perturbed_Hamiltonian}
\begin{split}
H'&= -2 \sum_i (X_i + Z_{i-1}X_{i} Z_{i+1}) \\
    &+(1+g)\sum_i(Z_{i}X_{i+1}Z_{i+2}X_{i+3} +X_{i}Z_{i+1}X_{i+2}Z_{i+3}) 
    \end{split}
\end{equation}
We also note that $H'$ cannot have a trivially-gapped phase due to the type-III mixed anomaly of the $\mathbb{Z}_2^3$ symmetry \cite{propitius1995topological,anomaly_wen_2015,yoshida_2016_symmetry}.

\begin{figure}[t!]
    \centering
    \includegraphics[width=1\linewidth]{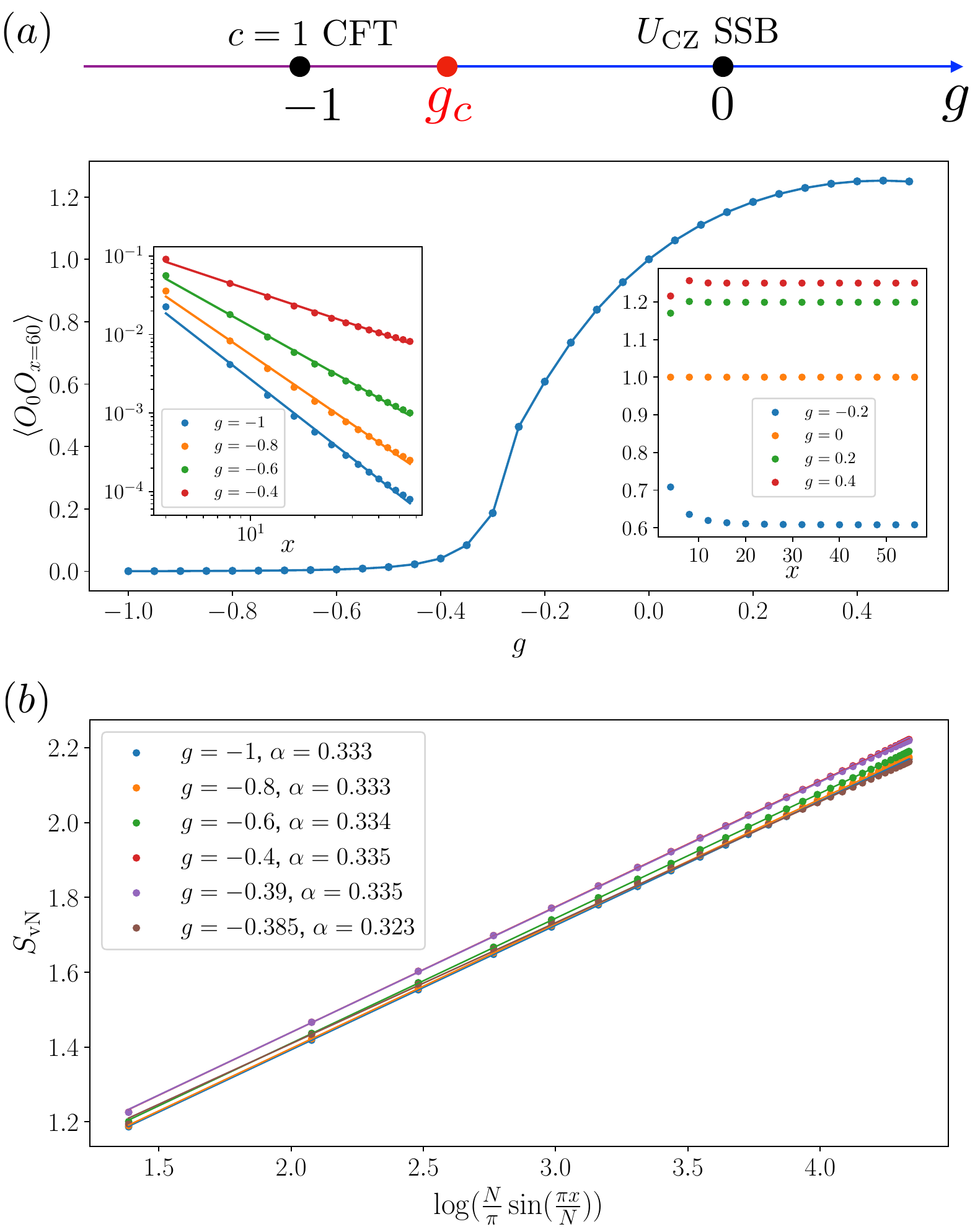}
    \caption{(a) The phase diagram of $H'$(Eq.\ref{eq:1d_perturbed_Hamiltonian}) and the correlations of non-invertible charged operators $\expval{O_0O_x}$ as a function of perturbation strength $g$ at a fixed $x=60$ and $N=120$. Right inset: in the $U_{\text{CZ}}$ SSB phase, $\expval{O_0O_x}$ saturates to a constant with the separation $x$. Left inset: in the CFT phase, $\expval{O_0O_x} \sim  x^{-\nu}$ with $\nu(g=-1) = 2.108$, $\nu(g=-0.8) = 1.858$, $\nu(g=-0.6) = 1.517$, and $\nu(g=-0.4) = 0.903$. (b) Entanglement entropy $S_{\text{vN}}$ in the CFT phase. $\alpha \equiv \frac{c}{3}\approx \frac{1}{3}$ with $c$ being the central charge. } 
    \label{fig:phase}
\end{figure}
Combining analytical arguments and the numerics with Density Matrix Renormalization Group (DMRG) \cite{1992_dmrg_white,1993_dmrg_white}, carried out using the ITensor library\cite{itensor}, we establish the phase diagram in Fig.\ref{fig:phase}(a). Below we discuss the essential features of the phase diagram with selected numerical results in Fig.\ref{fig:phase}. The remaining numerical results supporting the main text can be found in Appendix.\ref{appendix:parent_1d}. 

We find a critical point $g \approx -0.4$ that separates the gapped phase with $U_{\text{CZ}}$ spontaneous symmetry breaking (SSB) and the gapless phase described by a conformal field theory (CFT) with a central charge $c=1$. A representative of the former is at $g=0$, which gives the frustration-free gapped Hamiltonian (Eq.\ref{eq:1d_hamiltonian}) with two degenerate ground states $\ket{+}^{\otimes N}$ and $\ket{\text{cluster}}$. A representative of the latter is at $g=-1$, which gives the model ($\sum_i- X_i -Z_{i-1}X_{i}Z_{i+1}$), i.e. the critical point of a second-order transition between a trivial/non-trivial SPT under $\mathbb{Z}_2 \times \mathbb{Z}_2$ symmetry \cite{2009_Bartlett_cluster_transition,tsui2017phase,verresen2017one}.

In the $U_{CZ}$ SSB phase, we numerically find that the energy difference between the two lowest eigenstates decays with the system size $N$ (presumably vanishes as  $N\to \infty$) while the energy difference between the third eigenstate and the second eigenstate remains finite as increasing $N$. The order parameter $O_i O_j$ with $O_i= X_i - Z_{i-1}X_iZ_{i+1}$ saturates to a non-zero constant at long distances (Fig.\ref{fig:phase}(a)), providing a direct diagnosis of the $U_{\text{CZ}}$ SSB phase.

In the CFT phase, we numerically compute the bipartite entanglement entropy and find $c\approx 1$ by fitting the data into the scaling form $S_{\text{vN}} =  \frac{c}{3} \log\left[ \frac{N}{\pi}  \sin(\frac{\pi x}{N}  )   \right] + \text{const}$ \cite{wilczek_1994,calabrese_2004}, where $c$ is the central charge, and $x$ is the size of a subregion (Fig.\ref{fig:phase}(b)).  We also find power-law correlations among two distant local operators, which serves as another signal for the critical phase. In particular, this phase includes the point $g = -1$, described by the model $\sum_i - Z_{i}X_{i+1}Z_{i+2} - X_{i+1}$. It can be mapped to two decoupled critical transverse-field Ising chains via a KT transformation \cite{2023_KT_oshikawa}, implying its $c=1$ CFT nature. Furthermore, by a bosonization technique \cite{giamarchi2003quantum} (see Appendix.\ref{append:boson}), one finds the term $Z_{i}X_{i+1}Z_{i+2}X_{i+3} +X_{i}Z_{i+1}X_{i+2}Z_{i+3} $ is a marginal perturbation, which provides an analytical argument for the stability the CFT phase as well as the continuously varying critical exponents in the correlation functions. We also conjecture the critical point separating the critical phase and the $U_{CZ}$ SSB phase is the $SU(2)_1$ CFT. A detailed study of possible phase transitions out of this $U_{CZ}$ SSB phase is left for future work.

Finally, sending $g\to\infty$ gives the model $\sum_i(Z_{i}X_{i+1}Z_{i+2}X_{i+3} +X_{i}Z_{i+1}X_{i+2}Z_{i+3})$, whose frustration graph \cite{Planat_2008_graph,Flammia_2020_graph} is the same as that of the model $\sum_{i} Z_{i} Z_{i+1}Z_{i+2}Z_{i+3}  + \sum_i X_i$. The latter is known to be at a first-order critical point separating a disordered phase and a `modulated' phase \cite{1982_four_spin_Pfeuty,2014_four_spin_Florencio}. As a result, our model at $g\to \infty$ is at a first-order quantum critical point.

\section{2d 1-form non-onsite symmetry breaking - SPT soup}

\begin{figure}[t!]
    \centering
    \includegraphics[width=0.9\linewidth]{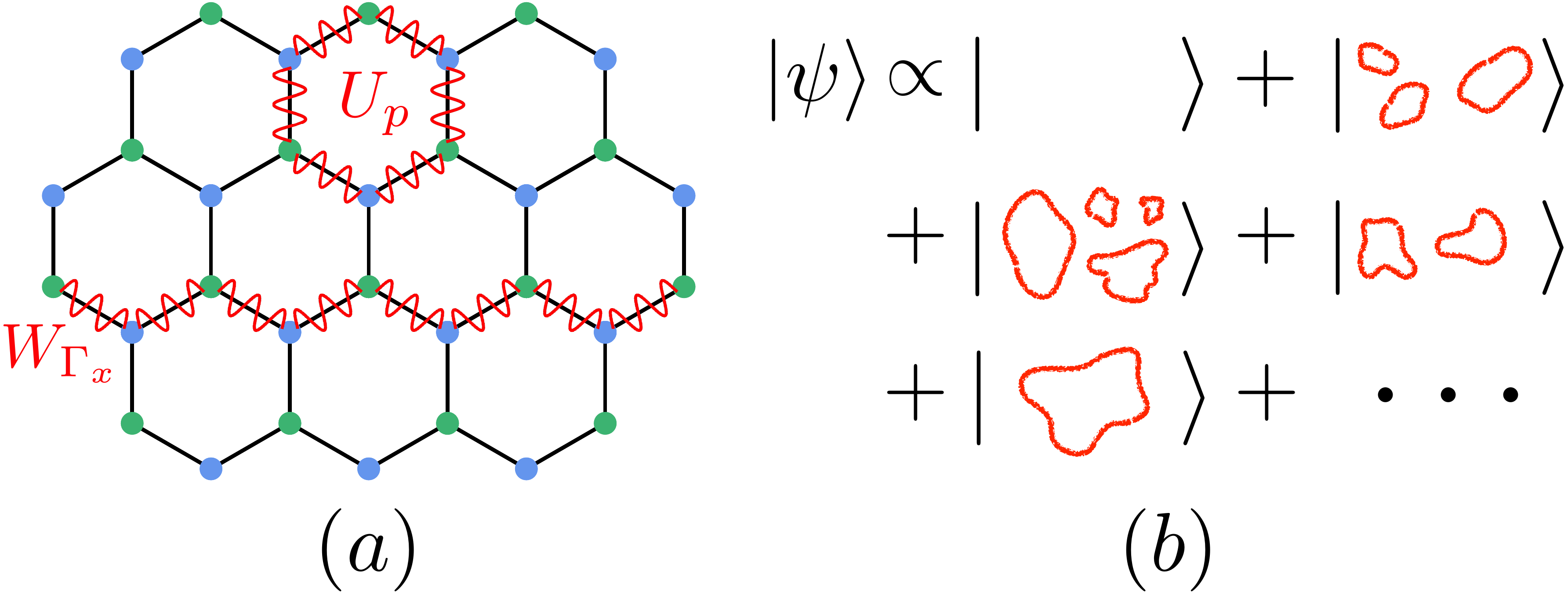}
    \caption{(a) 1-form non-onsite symmetry generators on a honeycomb lattice. (b) The SPT soup $\ket{\psi}$ is a condensate of $\mathbb{Z}_2\times \mathbb{Z}_2$ SPT cluster states on 1d loops.}  \label{fig:2d_soup}
\end{figure}

Now we extend the discussion to 2d and explore the physics of the spontaneous breaking of a 1-form non-onsite symmetry. We consider a honeycomb lattice with periodic boundary conditions and define qubits on vertices. We define the 1-form symmetry as $U= \prod_{\expval{ij} \in \text{loop} }\text{CZ}_{ij}$, i.e. the product of CZ gates acting along any closed loops, including both the contractible ones and the non-contractible ones. All the contractible 1-form symmetries can be obtained by taking the product of local plaquette symmetry:  $U_p= \prod_{\expval{ij} \in p  } \text{CZ}_{ij}$, i.e. the product of six CZ gates acting along the plaquette $p$ (see Fig.\ref{fig:2d_soup}(a)). 

To explore the 1-form non-onsite symmetry breaking, we first consider the following state:
\begin{equation} \label{eq:psi_soup}
\ket{\psi} \propto \prod_{p}  \left( 1+ U_p\right) \ket{+}, 
\end{equation}
where $\ket{+}$ denotes the product state in which every qubit is in the Pauli-X $+1$ eigenstate. $\ket{\psi}$ can be further expressed as $\sum_{\mathcal{C}} \prod_{\expval{ij} \in \mathcal{C}   } \text{CZ}_{ij}\ket{+} =  \sum_{\mathcal{C}}  \ket{\text{SPT}}_{\mathcal{C}} \ket{+}_{\overline{\mathcal{C}}}$.  $\ket{\text{SPT}}_{\mathcal{C}}$ denotes a $\mathbb{Z}_2 \times \mathbb{Z}_2$ cluster-state SPT along the contractible loop $\mathcal{C}$, and $\ket{+}_{\overline{C}}$ denotes the Pauli-X $+1$ product state for the qubits not on the loop $\mathcal{C}$. Namely, $\ket{\psi}$ is a superposition of loops of 1d SPT orders (see Fig.\ref{fig:2d_soup}(b)), and hence we name it SPT soup. Also, since the expectation value of 1d SPT entangler for any contractible loop $\mathcal{C}$, i.e. $\bra{\psi}\prod_{\expval{ij} \in \mathcal{C} }  \text{CZ}_{ij} \ket{\psi}$, is equal to one, $\ket{\psi}$ can be understood as an 1d SPT condensate in the 2d lattice.  

Our proposed SPT soup $\ket{\psi}$ also connects to the literature on SPT phases subject to measurements. Specifically, Ref.\cite{ashvin_2021_measurement,li_2023_measure_SPT} show that measuring a $\mathbb{Z}_2^3$ type-III fixed-point SPT \cite{propitius1995topological,yoshida_2016_symmetry} on a triangular lattice and post-selecting the measurement outcomes leads to a long-range entangled state, while the entanglement structure and the lattice wave function are not explored in detail. It turns out the measurement-induced state is exactly the SPT soup $\ket{\psi}$ that we propose above (see the derivation in Appendix.\ref{sec: SPT soup derivation} and Ref.\cite{vijay_Forthcoming}). With a similar calculation (Appendix.\ref{sec:levin_gu}), we also derive an SPT-soup-like wavefunction obtained by measuring one sublattice of the Levin-Gu SPT \cite{Levin_gu_spt_2012}, and argue the long-range entanglement based on its mixed anomaly between a 0-form (global) onsite symmetry and a 1-form non-onsite symmetry.

Now we discuss the signatures of long-range entanglement of the SPT soup from a few distinct perspectives. First, $\ket{\psi}$ cannot be prepared from a product state using finite-depth local unitary circuits. This is because one can construct three other orthogonal states (in the thermodynamic limit) that are locally indistinguishable on a torus, which indicates the long-range entanglement of these states via Lieb–Robinson bounds \cite{hastings_lrbound_2006}. Specifically, starting from the state $\ket{\psi}$ in  Eq.\ref{eq:psi_soup}, we can apply a product of $\text{CZ}$ gates along a non-contractable loop winding in $\hat{x}$ direction, denoted by $W_{\Gamma_x}$. This leads to $W_{\Gamma_x} \ket{\psi}  \propto \sum_{\mathcal{C}_x} \prod_{\expval{ij} \in \mathcal{C}_x   } \text{CZ}_{ij}\ket{+}$, where $\mathcal{C}_x$ is any closed loop winding along $\hat{x}$ direction. $\ket{\psi}$ and $W_{\Gamma_x} \ket{\psi} $ are expected to be orthogonal from each other in the thermodynamic limit (see Appendix.\ref{appendix:overlap}), but nevertheless, are locally indistinguishable: $\bra{\psi}W_{\Gamma_x} OW_{\Gamma_x} \ket{\psi}  =   \bra{\psi} O \ket{\psi} $ with $O$ being any local operator with a finite support. This is because one can always deform $W_{\Gamma_x}$ so that it does not overlap with $O$. Similarly, given $\ket{\psi}$,  one can apply $W_{\Gamma_y}$, i.e. a product of $\text{CZ}$ gates along a non-contractable loop winding in $\hat{y}$ direction. Therefore, $W_{\Gamma_x}$ and $W_{\Gamma_y}$ give rise to four locally indistinguishable ground states. This also implies that if their local parent Hamiltonian $H$ respects the 1-form non-onsite symmetries, these symmetries must be spontaneously broken at zero temperature.

While the local indistinguishability is a feature shared with the $Z_2$ toric-code topological order \cite{kitaev2003fault}, which exhibits a spontaneous 1-form onsite $\mathbb{Z}_2$ symmetry breaking \cite{nussinov2009sufficient,nussinov2009symmetry,higher_form_2015,Wen_higer_symmetries_2019,McGreevy_review_symmetry_2023}, the 1-form non-onsite symmetry broken states, i.e. SPT soup $\ket{\psi}$, do not admit a local, gapped parent Hamiltonian since there exist local operators whose two-point function decays algebraically \cite{2004_hasting_locality,2006_hasting_gap}. Specifically, by dividing the honeycomb lattice into two interlacing sublattices A and B colored in green and blue in Fig.\ref{fig:2d_soup}, we find $Z_{i_A}Z_{j_A}$ on the A sublattice (see Appendix.\ref{appendix:2point} for derivation): $\bra{\psi} Z_{i_A}Z_{j_A} \ket{\psi} = \frac{\sum_{\mathcal{C'}}    (\frac{1}{\sqrt{2}})^{\abs{\mathcal{C'}}} 2^{N_{\mathcal{C}'}}}{\sum_{\mathcal{C}} (\frac{1}{\sqrt{2}})^{\abs{\mathcal{C}}} 2^{N_{\mathcal{C}}}}$. Here $\mathcal{C}$ is the configuration of contractible loops, and $\mathcal{C'}$ is the configuration of contractible loops where there always exists a loop $\gamma$ connecting $i_A$ and $j_A$. The expression above is exactly the 2-leg watermelon correlator in the $O(2)$ loop model with the loop tension $K=\frac{1}{\sqrt{2}}$, which is at a critical point that separates a dilute phase and a dense phase \cite{Nienhuis_loop_1982,2019_lecture_loop}. At this critical point, the 2-leg watermelon correlator is known to decay algebraically \cite{duplantier1989two}:  $\bra{\psi} Z_{i_A}Z_{j_A} \ket{\psi} \sim \frac{1}{\abs{i_A - j_A}}$.  Since the SPT soup is symmetric under the $\mathbb{Z}_2$ symmetry $\prod_{i\in A} X_i $, namely $\prod_{i\in A} X_i \ket{\psi} = \ket{\psi}$, $\expval{Z_{i_A}}$ vanishes identically. This implies the algebraic decay of two-point connected correlation functions $\expval{Z_{i_A}Z_{j_A}} -   \expval{Z_{i_A}} \expval{Z_{j_A}}$, providing an alternative signature of long-range entanglement. 

Finally, we note that a parent Hamiltonian of the SPT soup state was found in Ref.\cite{li_2023_measure_SPT}. The proposed Hamiltonian, which is frustration-free, reads

\begin{equation}\label{eq:spt_soup_H}
H =  \sum_v (1-X_v) (1+Z_{v,1}Z_{v,2})(1+Z_{v,2}Z_{v,3}) - \sum_p U_p
\end{equation}
where $Z_{v,i}$ for $i=1,2,3$ are the three vertices neighboring to the vertex $v$; $U_p= \prod_{\expval{ij} \in p   } \text{CZ}_{ij}$ is the local constraint, which commutes with the first (vertex) term and energetically enforces $U_p=1$ in the ground states. The first (vertex) term of $H$ is a local projector, and it's simple to check that it annihilates the SPT soup $\ket{\psi}   \propto\prod_p (1+ U_p)  \ket{+}$, so $\ket{\psi}$ is indeed a ground state. Due to the algebraic correlations of the SPT soup, the parent Hamiltonian must be gapless. We leave the detailed analysis of such a Hamiltonian for future work.

\section{Summary and Discussion}

Recently there has been a surge of interest in using mid-circuit measurement and feedback unitary to efficiently prepare long-range entangled states  \cite{Raussendorf_2001_ghz,3d_cluster_state_2005,cirac_2008_optical,stace_2016_css,cirac_2021_locc,ashvin_2021_measurement,verresen2021_measurement_cold_atom,bravyi_2022_adaptive,lu2022measurement,ashvin_single_shot_2022,ashvin_hierarchy_2022,iqbal2023topological,foss2023experimental,Lu_mixed_feedback_2023,buhrman2023state,zhu2023nishimori,smith2023aklt,2023_set_wei,piroli2024approximating,wei_2024_simulation,iqbal2023_non_abelian,ibm_dynamic_circuit_2024,sahay2024finite,sahay2024classifying,adaptive_2024_mps_wei,stephen2024preparing}. Can we design a constant-depth measurement-feedback protocol to prepare the non-onsite symmetry breaking states, e.g. $\ket{+}^{\otimes N }+ \ket{\text{cluster}}$ in 1d and the SPT soup in 2d? In Appendix.\ref{appendix:preparation}, we present one such protocol that can prepare $ \ket{\psi_{\pm}}  \propto \ket{+}^{\otimes N} \pm  \ket{\text{cluster}}$ where the $+, -$ sign (determined by the measurement outcomes) is obtained with probability $\frac{1}{2}, \frac{1}{2}$ when $N\to  \infty$. This therefore present an experimentally feasible method to engineer and study the physics of non-onsite symmetry breaking on near-term quantum devices where mid-circuit measurement is available (e.g.  superconducting qubits or trapped-ion quantum computers) \cite{iqbal2023topological,foss2023experimental,ibm_dynamic_circuit_2024}.

Such a protocol can also be generalized to (probabilistically) prepare the superposition of any constant number of short-range entangled states, which may be of independent interest. Whether $\ket{\psi_+}$ can be deterministically prepared in constant depth remains an open question \footnote{Ref.\cite{stephen2024preparing} presented an alternative constant-depth adaptive protocol based on MPS (matrix product state) fusion, which can prepare the state $\ket{+}^{\otimes N}+ \ket{\text{cluster}}$ with (i) success probability 1  in 1d with \text{open} boundary condition and (ii) success probability $c=O(1)< 1$ in 1d with \text{periodic} boundary condition.}.

On the other hand, the constant-depth preparation of the 2d SPT soup $\prod_p (1+U_p) \ket{+}^{\otimes N}$ remains elusive. One naive attempt is to start from $\ket{+}^{\otimes N}$ and measure $U_p$ on every plaquette, in which case the pure-state trajectory corresponding to outcomes $U_p=1$ on all plaquettes is exactly the SPT soup. However, the measurement outcomes are generically random and need to be corrected. Since $\prod_p U_p=1$, the outcomes $U_p=-1$ will come in pairs, which nevertheless cannot be paired up with depth-1 unitary circuits due to the non-zero trace of $U_p$. This is in contrast to the measurement-based preparation for the $\mathbb{Z}_2$ toric code, where the pairs of defects (plaquette stabilizers with the $-1$ measurement outcome) can be annihilated with a depth-1 unitary string operator. The SPT soup therefore presents an intriguing example where `abelian $\mathbb{Z}_2$ defects' \footnote{$U_p=-1$ is regarded as an abelian $\mathbb{Z}_2$ defect since $U_p^2=1$.} cannot be efficiently annihilated.

The physics of the SPT soup and its parent Hamiltonian (Eq.\ref{eq:spt_soup_H}) also deserves further investigation. In particular, we do not have a clear understanding of the connection between the $\mathbb{Z}_2$ 1-form non-onsite symmetry and the criticality exhibited by the SPT soup. Could there be other symmetries and physical mechanisms that enforce the gaplessness in this model? Relatedly, while the SPT soup can be constructed on any lattice, it can be mapped to the $O(2)$ loop model with a string tension $\frac{1}{\sqrt{2}}$ only when the lattice is trivalent. When the lattice is not trivalent, the loops can branch and intersect, in which case, the corresponding statistical mechanics model is no longer an O(2) loop model, and the physics may depend on the details of the lattice. It would be interesting to explore the properties of SPT soups on various lattices.


Finally, it would be interesting to explore non-onsite SSB from the perspective of Topological Holography / Symmetry Topological Field Theory (SymTFT), a framework aiming to classify quantum phases from a topological order in one higher space dimension  \cite{ji2020categorical,kong2020algebraic,gaiotto2021orbifold,lichtman2021bulk,freed2024topological,moradi2023topological,bhardwaj2024categorical,bhardwaj2025gapped,huang2025topological,chatterjee2023symmetry}. We present a preliminary analysis with the SymTFT perspective in Appendix.\ref{append:sym_tft}.

\textbf{Note added}: During the completion of this manuscript, we became aware of an independent, forthcoming work \cite{vijay_Forthcoming}, which studies the gapless correlations that can arise in SPT states after measurements, and also investigates the wavefunction (Eq.\ref{eq:psi_soup}) and its algebraic correlations in this context.

\textbf{Acknowledgements}:
We thank Yimu Bao, Sarang Gopalakrishnan, Alexey Gorshkov, Alex Jacoby, Ali Lavasani, Yaodong Li, Mikhail Litvinov, Jacob Lin, Ruochen Ma, Shinsei Ryu, Pok Man Tam, Tzu-Chieh Wei, Zhi-Yuan Wei, Yichen Xu, and Yifan Frank Zhang for inspiring discussions. We also thank Sahand Seifnashri, Sagar Vijay, and Yichen Xu for valuable feedback on the manuscript. T.-C.L acknowledges the support of the RQS postdoctoral fellowship through the National Science Foundation (QLCI grant OMA-2120757). This work was also supported by the U.S. National Science Foundation under Grant No. NSF DMR-2316598 (YL). Use was made of computational facilities purchased with funds from the National Science Foundation (CNS-1725797) and administered by the Center for Scientific Computing (CSC). The CSC is supported by the California NanoSystems Institute and the Materials Research Science and Engineering Center (MRSEC; NSF DMR 2308708) at UC Santa Barbara.

\clearpage
\widetext

\appendix

\section{General remark on non-onsite symmetry}\label{1d:remark}

\subsection{Non-onsite to onsite symmetries via duality} 
Here we show that the intrinsically non-onsite symmetry $U_{\text{CZ}}=\prod_i  CZ_{i,i+1 }$ can become an onsite symmetry under a Kramer-Wannier duality. To start, we consider a 1d lattice of even number of vertices, and write $U_{\text{CZ}}$ as 
\begin{equation}
    U_{CZ}=\prod_{i}CZ_{2i-2,2i-1}CZ_{2i-1,2i}=\prod_{i}\frac{1+Z_{2i-1}+Z_{2i-2}Z_{2i}-Z_{2i-2}Z_{2i-1}Z_{2i}}{2},
\end{equation} 

A Kramers-Wannier duality on the even sublattice gives the following mapping for local operators symmetric under $\prod_{i}X_{2i}$: 
\begin{equation}
    X_{2i}\rightarrow Z_{2i}Z_{2i+2},\quad Z_{2i}Z_{2i+2}\rightarrow X_{2i+2}.
\end{equation}
As a result, 
\begin{equation}
U_{\text{CZ}} \to U'=\prod_{i}CX_{2i-1,2i}, 
\end{equation}
and $U'$ is onsite by grouping the adjacent qubits on vertex $2i-1$ and vertex $2i$.

There could also exist non-onsite symmetries that cannot be brought onsite even under a duality transformation. We expect the anomalous $\mathbb{Z}_2$ symmetry \cite{chen2011two} generated by $U_{CZX}=\prod_i CZ_{i,i+1}\prod_i X_i$ in 1d is one such example, as we argue below based on SymTFT description (see Sec.\ref{append:sym_tft} for more discussion). Within this description, this anomalous symmetry corresponds to a semionic anyon line in the 2d bulk topological order. A duality transformation in a 1d system corresponds to a change of reference gapped boundary. No matter what duality transformation we apply, the dual symmetry of $U_{CZX}$ is always given by the same semionic anyon line from the bulk, hence also being anomalous. This suggests no duality can bring the anomalous $U_{CZX}$ operator onsite.

To conclude, we expect that there exist three hierarchies for non-onsite symmetries. The first hierarchy includes the symmetry operators that can be made onsite by a conjugation, i.e. there exists a unitary operator $V$, such that the symmetry operator $U=V(\prod_s U_s) V^{\dagger}$. The second and the third hierarchies are intrinsically non-onsite symmetries, where the former includes the operators that can be brought onsite by some duality transformations, and the latter cannot be brought onsite even under dualities. The non-onsite $U_{\text{CZ}}$ symmetry studied here is a representative of the second hierarchy. 

\subsection{Equivalence between onsite and non-onsite symmetries with ancillae}\label{append:ancilla}

Let us consider introducing an ancilla qubit at state $\ket{+}_{\tau}$ on each even site of the spin chain. There is a global symmetry $\prod_i \tau^x_{2i}$ on the ancilla state. With this $\mathbb{Z}_2$ symmetry and the non-onsite symmetry $U_{\text{CZ}}$, one can define a diagonal non-onsite symmetry $\mathbb{Z}_2^{d}$: $U_d  =\prod_i CZ_{i,i+1}\prod_j \tau^x_{2j}$. After the conjugation of the finite-depth local unitary circuit
\begin{equation}
    V=\prod_i \frac{1}{2}\Big(1+\tau^z_{2i}+CZ_{2i-1,2i}CZ_{2i,2i+1}-\tau^z_{2i}CZ_{2i-1,2i}CZ_{2i,2i+1}\Big),
    \label{eq:conjugation_V}
\end{equation}
$U_d$ is mapped to the onsite symmetry $U'_{d}=\prod_i \tau^x_{2i}$. Therefore, we see that \emph{non-onsite-ness is not stable}, in that a non-onsite $U_{CZ}$ symmetry in the second hierarchy becomes intrinsically onsite, after including ancillae and taking the diagonal symmetry.

We also note that if considering the two additional symmetries $Z_2^e$ and $Z_2^o$ generated by onsite unitaries $U_e=\prod_i X_{2i}, U_o=\prod_i X_{2i+1}$, which together with $U_{\text{CZ}}$ forms a type-III mixed anomaly \cite{propitius1995topological,anomaly_wen_2015}, under the conjugation of $V$,  $U_e$ and $U_0$ become

\begin{equation}
\begin{split}
    &U'_e=\prod_i X_{2i}\cdot \frac{1}{2}\Big(1+\tau^z_{2i}+Z_{2i-1}Z_{2i+1}-\tau^z_{2i}Z_{2i-1}Z_{2i+1}\Big),\\ 
    &U'_o=\prod_i X_{2i+1},\\
    \end{split}
\end{equation} 
where the $\mathbb{Z}_2^e$ symmetry becomes non-onsite. In fact, due to the anomaly, there is always at least one non-onsite symmetry operator no matter how we make conjugation.

\subsection{Onsite symmetry can also have non-invertible charge operators}\label{append:remark_charged_op}

In the main text, we show that any unitary symmetry with a non-zero trace cannot have invertible charged operators. The charged operator $X_i - Z_{i-1}X_iZ_{i+1}$ of the non-onsite symmetry $U_{\text{CZ}} = \prod_i \text{CZ}_{i,i+1}$ is one such example. However, some onsite symmetries can also have a non-zero trace, hence having a non-invertible charged operator. For instance, consider a 1d lattice of qutrit, with a local Hilbert space spanned by 3 computational bases $\ket{0}, \ket{1}, \ket{2}$, one defines a global $\mathbb{Z}_2$ onsite symmetry $U=  \prod_i (\ket{0} \bra{1}  + \ket{1} \bra{0}+ \ket{2} \bra{2 } )_i $, which exchanges between $\ket{0}$ and $\ket{1}$ on every site while acting trivially on $\ket{2}$. $U$ is onsite but due to its non-zero trace, the charged operator (i.e. $O_i = \ket{0}\bra{0}  -  \ket{1}\bra{1}  $  ) must be non-invertible.

\section{Details on 1d non-onsite symmetry breaking}

\subsection{Proof for intrinsically non-onsite $\prod_i CZ_{i,i+1}$ symmetry}\label{sec:prove_non_onsite}
Here we show that $U_{CZ}=\prod_i CZ_{i,i+1}$ is intrinsically non-onsite by a proof of contradiction. To start, we calculate the trace of $U_{CZ}$ on a length-$N$ chain: 
\begin{equation}
    \tr U_{CZ}=2^N\bra{+}^{\otimes N} U_{CZ}\ket{+}^{\otimes N }=2^N\cdot \bra{+}^{\otimes N }\ket{\text{cluster}}=2\cdot2^{\frac{N}{2}},
    \label{eq:CZtrace}
\end{equation}
where we have used $\bra{+}^{\otimes N }\ket{\text{cluster}} = 2^{1- \frac{N}{2}}$.  We assume there is a unitary conjugation that can bring $U_{CZ}$ to a translational invariant onsite operator $\bigotimes_{s=1}^{N/m}u$, where $u$ is a local unitary acting on each site that contains $m$ neighboring qubits. Since a matrix trace is invariant under unitary conjugation, one should have 

\begin{equation}
    \tr U_{CZ}= \tr \left(\bigotimes_{s=1}^{N/m}u\right).
\end{equation}

To proceed, we notice that for any two unitary operators $U_A$ and $U_B$ acting on the Hilbert space $\mathcal{H}_A, \mathcal{H}_B$, the trace of their tensor product over the Hilbert space $\mathcal{H}_{AB}= \mathcal{H}_A\otimes \mathcal{H}_B$ satisfies 
\begin{equation}
    \tr_{AB}(U_A\otimes U_B)= (\tr_A U_A ) (\tr_B U_B).
\end{equation}
This implies 

\begin{equation}
    \tr U_{CZ}= (\tr_s u)^{N/m}= k^{N/m}
        \label{eq:traceequal}
\end{equation}
where $k\equiv \tr_s u$ is the trace of $u$ on the $2^{m}$-dim local Hilbert space. In particular, since $U_{CZ}$ generates a $\mathbb{Z}_2$ symmetry, the eigenvalues of $u_s$ are $\pm 1$, which implies that the trace $k$ must be an integer. On the other hand, comparing Eq.\ref{eq:CZtrace}, Eq.\ref{eq:traceequal}, one finds
\begin{equation}
    k=2^{\frac{m}{2}+\frac{m}{N}}, 
\end{equation}
which is an integer only when $m=N$. This contradicts the assumption that $U_{\text{CZ}}$ can be brought onsite by a unitary conjugation and a regrouping of constant $m$ neighboring qubits into a single site. Therefore, we conclude that $U_{CZ}$ is intrinsically non-onsite.

\subsection{Proof of the two-fold ground-state degeneracy}\label{appendix: proof of GSD}
Here we prove that the frustration-free Hamiltonian 

\begin{equation}\label{eq:1d_hamiltonian_appendix}
H = \sum_{i} (1 - X_i) (1 - Z_{i+1}X_{i+2} Z_{i+3}) + \sum_i (1 - Z_i X_{i+1} Z_{i+2}) (1 - X_{i+3})
\end{equation}
only has two ground states, i.e. $\ket{+}^{\otimes N}$ and $\ket{\text{cluster}}$. Our proof strategy is inspired by the proof technique presented in Ref.\cite{duality_Huang_2024}, which proves (i) the ground state degeneracy of O'Brien-Fendley Hamiltonian \cite{o2018lattice} with order-disorder coexistence in one space dimension and (ii) the ground state degeneracy of a frustration-free Hamiltonian with the coexistence between a product state and toric-code ground states in three space dimensions.

First, our Hamiltonian respects the two $Z_2$ onsite global symmetries given by $U_{o}=\prod_{i \in  odd} X_i, U_{e}=\prod_{i \in  even} X_i$. This indicates $H$ and these two symmetry generators share the same eigenstates, so the ground states of $H$ can be divided into four subspaces labeled by the eigenvalues of $U_o, U_e$: $(1,1), (1,-1), (-1,1), (-1,-1) $. We will first focus on the sector  $(U_o, U_e)=  (1,1)$. Later we will see that the other three sectors contain no ground states.

Now we show in the subspace with $U_o= U_e =1$, $\ket{+}^{\otimes N}$ and $\ket{\text{cluster}}$ are the only two ground states. To begin with, since $H$ is a sum of local projectors, any ground states $\ket{\psi}$ must be annihilated by each local term: 

\begin{equation}\label{append:1d_pq_relation}
P_i Q_{i+2} \ket{\psi}  =Q_{i}P_{i+2} \ket{\psi}   =0
\end{equation}
with 
\begin{equation}
\begin{split}
&P_i \equiv 1-X_i,  \\
&Q_i\equiv 1-Z_{i-1} X_i Z_{i+1}
\end{split}
\end{equation}
In the subspace with $U_o= U_e =1$, any states can be expanded in the $X$ basis as 

\begin{equation}
\ket{\psi}  =  \sum'_{\sigma \in \{ +, -\}^N }  \psi(\sigma) \ket{\sigma}
\end{equation}
where $\sum'_{\sigma \in \{ +, -\}^N } $ denotes a restricted sum over $\sigma$ with $\prod_{i\in odd}\sigma_i = \prod_{i\in even}\sigma_i = 1$. 

Eq.\ref{append:1d_pq_relation} provides a relation among amplitudes $\psi(\sigma)$ of various basis $\ket{\sigma}$. To see this, we first consider the action of $P_iQ_{i+2}$ on a product state $\ket{\sigma}$

\begin{equation}
P_iQ_{i+2} \ket{...\sigma_i \sigma_{i+1} \sigma_{i+2} \sigma_{i+3} ...} = \begin{cases}    
 0  \quad\qquad\qquad\qquad\qquad\qquad\qquad\qquad\qquad\qquad\qquad \text{for } \sigma_i =+ \\
2 \left[\ket{...\sigma_i \sigma_{i+1} \sigma_{i+2} \sigma_{i+3} ...}  -  \sigma_{i+2}\ket{...\sigma_i \overline{\sigma}_{i+1} \sigma_{i+2} \overline{\sigma}_{i+3} ...} \right] \quad  \text{for } \sigma_i =-     
\end{cases}
\end{equation}
with $\overline{\sigma}_i  \equiv  - \sigma_i$. Therefore the constraint $P_iQ_{i+2} \ket{\psi} =0$  implies  

\begin{equation}
\psi(...\sigma_i \sigma_{i+1} \sigma_{i+2} \sigma_{i+3} ...)  =  \sigma_{i+2} \psi(...\sigma_i \overline{\sigma}_{i+1} \sigma_{i+2} \overline{\sigma}_{i+3} ...)   \text{   for    } \sigma_i =-.
\end{equation}

Similarly,  the constraint $Q_{i}P_{i+2} \ket{\psi} =0$ implies 

\begin{equation}
\psi(...\sigma_{i-1} \sigma_{i} \sigma_{i+1} \sigma_{i+2} ...)  =  \sigma_{i}  \psi(...\overline{\sigma}_{i-1} \sigma_{i} \overline{\sigma}_{i+1} \sigma_{i+2} ...)   \text{   for    } \sigma_{i+2} =-.
\end{equation}

The above two relations can be summarized as follows:\\
\noindent \textbf{Rule 1}: Given a configuration $\ket{\sigma}\neq \ket{+}^{\otimes N}$, if  $\sigma_i=-$, one can flip  $\sigma_{i+1},\sigma_{i+3}$ to obtain the configuration $\sigma'$ with $\psi (\sigma) = \sigma_{i+2}\psi(\sigma')$, or flip $\sigma_{i-1},\sigma_{i-3}$ to obtain the configuration $\sigma'$ with $\psi(\sigma) = \sigma_{i-2} \psi(\sigma')$.\\

Rule 1 further implies the following: 

\noindent \textbf{Rule 2:} When $\sigma_i \neq \sigma_{i+2}$, one has the relation between the amplitudes

\begin{equation}
\psi(...  + \sigma_i - ...)    =  \sigma_{i}  \psi(... - \sigma_i +...  )
\end{equation}
by exchanging $\sigma_i$ and $  \sigma_{i+2}$ without affecting other signs, i.e. 

\begin{equation}\label{append:fact}
...  - \sigma_i + ...  ~\longleftrightarrow ~...  + \sigma_i - ... 
\end{equation}

To see this, without loss of generality, let's consider the consecutive five spin configurations: $\sigma_1- \sigma_3 +   \sigma_5$, and there are four choices of $(\sigma_1,\sigma_5)$:  \\ 

\noindent (1) $(\sigma_1,\sigma_5) = (- ,-):\quad $ 
$-  + \sigma_3  -  -  ~\overset{(2,4)}{\longleftrightarrow} ~ - - \sigma_3 + -$.

\vspace{2mm}
\noindent (2) $(\sigma_1,\sigma_5) = (- ,+):\quad$ 
$-  +\sigma_3  -  +  ~\overset{(2,4)}{\longleftrightarrow} ~ - - \sigma_3 + +$. 

\vspace{2mm}

\noindent (3) $(\sigma_1,\sigma_5) = (+ ,-):\quad$  
$+  + \sigma_3  -  -  ~\overset{(2,4)}{\longleftrightarrow} ~ + - \sigma_3 + -$. 
\vspace{2mm}

\noindent (4) $(\sigma_1,\sigma_5) = (+ ,+):\quad$    $   +  + \sigma_3  -  +  ~\overset{(1,3)}{\longleftrightarrow} ~  -  + \overline{\sigma}_3  -  +  $\\

$\qquad \qquad\qquad\qquad \qquad\qquad\qquad ~~~\, \overset{(2,4)}{\longleftrightarrow} ~   -  - \overline{\sigma}_3  +  +  $\\ 

$\qquad \qquad\qquad\qquad \qquad\qquad\qquad ~~~\,\overset{(3,5)}{\longleftrightarrow} ~ -  - \sigma_3 + -    $\\

$\qquad \qquad\qquad\qquad \qquad\qquad\qquad ~~~\,\overset{(2,4)}{\longleftrightarrow} ~ -  + \sigma_3 - -  $\\

$ \qquad \qquad\qquad\qquad \qquad\qquad\qquad ~~~\,\overset{(1,3)}{\longleftrightarrow} ~ + + \overline{\sigma}_3- - $\\ 

$\qquad \qquad\qquad\qquad \qquad\qquad\qquad ~~~\,\overset{(2,4)}{\longleftrightarrow} ~  + - \overline{\sigma}_3 + - $\\

$ \qquad \qquad\qquad\qquad \qquad\qquad\qquad ~~~\,\overset{(3,5)}{\longleftrightarrow} ~  + - \sigma_3 + +  $  \\
where the numbers (e.g. 2,4) right above the left-right arrows denote the indices for the flipped spins. Combining the results of the above four cases proves Rule 2.

One application of Rule 2 is that for any $\sigma\neq +++...$,  one can move all the $-$ signs on odd (even) sites into contiguous subregions on the odd (even) sublattice, e.g.

\begin{equation}
\includegraphics[scale=0.2]{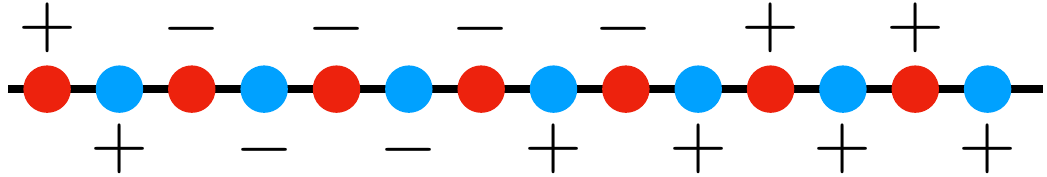},  
\end{equation}
where the spins not shown in the figures are all in the $+$ sign. We note that the number of $-$ signs on the odd (even) sublattice must be even due to the constraint $\prod_{i\in odd}\sigma_i = \prod_{i\in even}\sigma_i =1$. Combining Rule 1 and Rule 2, all the $-$ signs on the odd sublattice can be annihilated, and the number of $-$ signs on the even sublattice will be reduced to two, i.e. 
\begin{equation}
\includegraphics[scale=0.2]{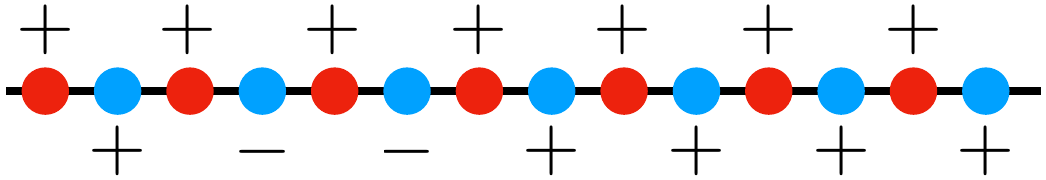}. 
\end{equation}

The analysis above shows that all the X-basis states $\ket{\sigma} $ with $\prod_{i\in odd} \sigma_i =\prod_{i\in even} \sigma_i =1$ that are not equal to  $  \ket{+}^{\otimes N }$  can all be connected. In other words, in this symmetric subspace, in addition to $\ket{+}^{\otimes N}$, there can only be one extra ground state, which must be $\ket{\text{cluster}}$.

\vspace{2mm}

\noindent \textbf{Proving the ground states of $H$ must only exist in the symmetric subspace $U_e = U_o =1$}:

Here we prove that the ground states of $H$ must only exist within the subspace with $U_e = U_o =1$, thereby establishing the two-fold ground-state degeneracy of $H$. To prove this, we will first express a general ground state on the $X$ basis

\begin{equation}
\ket{\phi} = \sum_\sigma  \phi(\sigma) \ket{\sigma }. 
\end{equation}
and show that the wave amplitude $\phi(\sigma)$ is exactly zero whenever the condition $\prod_{i\in odd}  \sigma_i =  \prod_{i\in even} \sigma_i =1 $ is not satisfied. There are only three classes of the configurations we need to consider: \\

\noindent (1) ($\prod_{i\in odd} \sigma_i,\prod_{i\in even} \sigma_i)= (1,-1)$: 
all the $\sigma$ configurations can be connected using Rule 1 and Rule 2 above, meaning the corresponding amplitudes $\phi(\sigma)$ are all the same up to a $\pm$ sign. In particular, one finds $\phi(\sigma)=0$ by considering the following: starting with $\sigma$ with a single $-$ on an odd sublattice with $\phi(...+-+++...)$, one has

\begin{equation}
\begin{split}
&\phi(...+-+++...) \overset{R_1}{=} \phi(...+--+-...)  \overset{R_2}{=} - \phi(...++---...) \\
&\overset{R_2}{=} - \phi(...-+--+...) \overset{R_1}{=} - \phi(...+++-+...) \overset{R_1}{=}  - \phi(...+-+++...)  
\end{split}
\end{equation}
where the plotted signs are for $\sigma_i$ with $i=1,2,3,4,5$, and $R_1, R_2$ above the equal signs indicate the use of Rule 1 or Rule 2. Also, note that the third equal sign is accomplished by sequentially moving the $-$ sign on $\sigma_5$ towards the right until stopping at $\sigma_1$. The above shows that the amplitude $\phi=0$ for the X basis states in this class. \\

\noindent (2) ($\prod_{i\in odd} \sigma_i,\prod_{i\in even} \sigma_i)= (-1,1)$: using the same analysis as above, $\phi=0$ for any $\sigma$ configurations in this subspace.\\

\noindent (3) ($\prod_{i\in odd} \sigma_i,\prod_{i\in even} \sigma_i)= (-1,-1)$: using the Rules above, all the configurations in this class can be connected to $... +--++...$ with only two neighboring $-$ signs on $\sigma_2, \sigma_3$. Keep applying Rule 2, one can move the minus signs from $\sigma_2$ to $\sigma_4$, $\sigma_6$, all the way back to $\sigma_2$, with a minus sign in the amplitude: $\phi(... +--++...) = -  \phi(... +--++...) $, which implying $\phi=0$. \\

To conclude, the ground states of $H$ only exist in the subspace with $U_o= U_e=1$, so $\ket{+}^{\otimes }$ and $\ket{\text{cluster}}$ are the only two ground states of $H$.

\vspace{2mm}

\subsection{Proof of a finite energy gap}\label{appendix:gap}
In this section, we prove there is a finite energy gap above the ground state subspace of the frustration-free Hamiltonian in \eqref{eq:1d_hamiltonian_appendix}. Our strategy is largely inspired by the one used in \cite{duality_Huang_2024}.\newline
\indent We assume that the system size $L$ is a multiple of an integer $n \geq 2$. We coarse-grain the lattice by a factor of $n$ so that each new site, labelled by $L = 1,..., \frac{L}{n}$, contains the n sites $i = n(I - 1) + 1, ..., nI$. Consider the local Hamiltonian restricted to $I$ and $I + 1$
\begin{equation}
    h_{I, I+1} \coloneq \sum_{i = n(I - 1) + 1}^{n(I + 1) - 2} \left(P_{i}Q_{i+1} + Q_{i}P_{i+3} \right),
\end{equation}
where $P_i = \frac{1 - X_i}{2}$ and $Q_{i} = \frac{1 - Z_i X_{i+1}Z_{i+2}}{2}$.
\newline
\indent Following the same proof in \cite{duality_Huang_2024}, one can lower bound $H$ by another Hamiltonian $\bar{H} \coloneq \sum_{I = 1}^{L / n} \Pi_{I, I+1}^{\perp}$, where $\Pi_{I, I+1}$ is the projector onto the (nontrivial) kernel of $h_{I,I+1}$. More precisely, one can prove that 
\begin{equation}
    H \geq \epsilon \bar{H}
\end{equation}
where $\epsilon > 0$ is the smallest positive eigenvalue of $h_{I,I+1}$ and $\epsilon$ is independent of $I$ due to translation invariance but can depend on the size of new site $n$. Due to frustration-freeness, $H$ has the same groundstates as $\bar{H}$, so proving the finite energy gap of $\bar{H}$ implies $H$ also has a finite energy gap. Let's denote the finite energy gap of $H$ as $\Delta$ and that of $\bar{H}$ as $\bar{\Delta}$. \newline
\indent We use the Knabe's argument\cite{knabe1988energy} to prove the finite energy gap of $\bar{H}$. We want to prove that $\bar{H}^2 \geq \Delta_0 \bar{H}$. Since $\bar{H}$ is positive semi-definite, the above inequality implies that $\bar{\Delta} \geq \Delta_0$. Following \cite{duality_Huang_2024}, one can arrive at the following expression of $\Delta_0$
\begin{equation}
    \Delta_0 = 1 - 2\delta_n,
\end{equation}
where $\delta_n$ is defined as $\delta_n \coloneq || \Pi_{I, I+1} \Pi_{I+1,I+2} - \Pi_{I,I+1} \wedge \Pi_{I+1, I+2} ||$, where $|| \cdot ||$ is the operator norm and $\wedge$ is the orthogonal projection.\newline
\indent Thus, the gap $\Delta$ of Hamiltonian $H$ in \eqref{eq:1d_hamiltonian_appendix} satisfies 
\begin{equation}
    \Delta \geq 1 - 2\delta_n.
\end{equation}
We shall derive an upper bound of $\delta_n$ below, where many calculation details differ from those in \cite{duality_Huang_2024}. \\

\noindent \textbf{Upper bound on $\delta(A,B)$}:\\
 We consider restricting the Hamiltonian to a local connected region $\Lambda$,
\begin{equation}
    h_{\Lambda} \coloneq \sum_{i \in \Lambda: i+3 \in \Lambda} \left(P_{i} Q_{i+1} + Q_{i}P_{i+3} \right),
\end{equation}
where $P_i \coloneq \frac{1 - X_i}{2}$ and $Q_i \coloneq \frac{1 - Z_i X_{i+1} Z_{i+2}}{2}$ are projectors. $h_{\Lambda}$ is the sum of terms that are entirely supported in region $\Lambda$. Using the argument presented in \ref{appendix: proof of GSD}, we can show that for $|\Lambda| \geq 4$ and, for simplicity, $|\Lambda|$ is even, $h_{\Lambda}$ has five ground states,
\begin{equation}
    \begin{split}
        &\ket{+}_{\Lambda} \coloneq \ket{+++...+}_{\Lambda},~~~\ket{(1,1)}_{\Lambda} \coloneq \prod_{i \in \Lambda: i+1\in \Lambda}CZ_{i,i+1}\ket{+++...+}_{\Lambda},~~~\ket{(-1,1)}_{\Lambda} \coloneq \prod_{i \in \Lambda: i+1\in \Lambda}CZ_{i,i+1}\ket{-++...+}_{\Lambda} \\
        &~~~~~~~~~~~~~~~~~\ket{(1,-1)}_{\Lambda} \coloneq \prod_{i \in \Lambda: i+1\in \Lambda}CZ_{i,i+1}\ket{+-+...+}_{\Lambda},~~~\ket{(-1,-1)}_{\Lambda} \coloneq \prod_{i \in \Lambda: i+1\in \Lambda}CZ_{i,i+1}\ket{--+...+}_{\Lambda},
    \end{split}
\end{equation}
where $\ket{(m,n)}_{\Lambda}$ ($m,n = \pm 1$) are nothing but cluster state defined on an open interval $\Lambda$ in symmetry sector $(m,n)$.

The five ground states have the following inner product relations:
\begin{equation}
    \prescript{}{\Lambda}{\braket{+}{(m,n)}_{\Lambda}} = \frac{(-1)^{mn}}{2^{\abs{\Lambda} /2 }},~~~
\prescript{}{\Lambda}{\braket{(m',n')}{(m,n)}_{\Lambda}} = \delta_{m',m}\delta_{n',n}.
\end{equation}

\indent Consider two overlapping and connected intervals $A$ and $B$ such that $\abs{A} \geq 8$, $\abs{B} \geq 8$, and $\abs{A \cap B} \geq 4$. Let $A' \coloneq A \setminus B$ and $B' \coloneq B \setminus A$. Note that $A'$, $A \cap B$, and $B'$ are also connected intervals. Without loss of generality, we can assume the following geometry for $A$ and $B$
\begin{equation*}
    \includegraphics[scale=0.2]{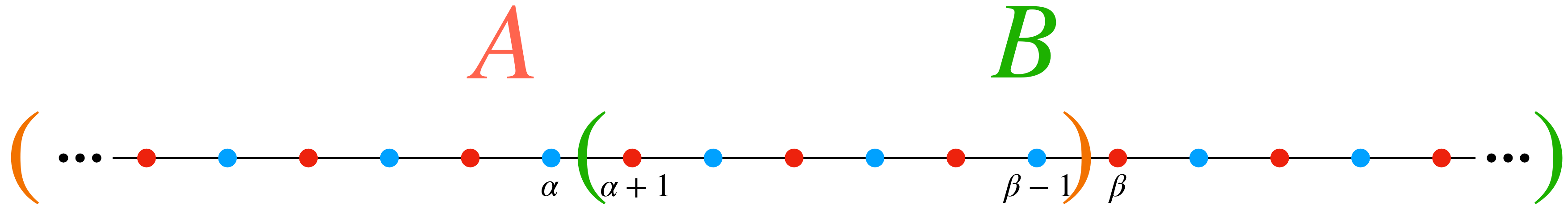}
\end{equation*}$m$ labels the symmetry charge on the red sublattice and $n$ labels the symmetry charge on the blue sublattice. The labeling of the boundary qubits will be used later in the proof.\newline
\indent Before deriving the overlap between ground states in $A$ and $B$, we want to emphasize a trick that is going to be extensively used in the rest of the proof. We notice that$\ket{(m,n)}_A$ can be decomposed as 
\begin{equation}
    \ket{(m,n)}_A = CZ_{\alpha, \alpha+1} \ket{(m,n)}_{A'} \ket{1,1}_{A \cap B} = CZ_{\alpha, \alpha + 1} \ket{(1,1)}_{A'}\ket{(m,n)}_{A \cap B},
\end{equation}
where intuitively it means that the information of the symmetry charge sectors $(m,n)$ can be put into either region $A'$ or region $A \cap B$.
Then, we have
\begin{equation}
    \begin{split}
        &\prescript{}{A}{\braket{+}{+}_B} = \ket{+}_{B'}  \prescript{}{A'}{\bra{+}},\\
        \\
        & \prescript{}{A}{\braket{(m,n)}{+}_B} =  \prescript{}{A'}{\bra{(m,n)}} \prescript{}{A \cap B}{\bra{(1,1)}} CZ_{\alpha,\alpha+1}  \ket{+}_{A\cap B} \ket{+}_{B'} \\
        &~~~~~~~~~~~~~~~~~~=\frac{1}{2^{\abs{A \cap B} / 2}} \ket{+}_{B'} \prescript{}{A'}{\bra{(m,n)}}\\
        \\
        &\prescript{}{A}{\braket{(m,n)}{(m',n'))}_B} = \prescript{}{A'}{\bra{(m,n)}} \prescript{}{A \cap B}{\bra{(1,1)}} CZ_{\alpha,\alpha+1}  CZ_{\beta - 1,\beta}\ket{(1,1)}_{A\cap B} \ket{(m',n')}_{B'} \\
        &~~~~~~~~~~~~~~~~~~~~~~~~~~~= \frac{1}{2} \ket{(m',n')}_{B'}  \prescript{}{A'}{\bra{(m,n))}} + \frac{1}{2} \ket{(\bar{m}',n')}_{B'}  \prescript{}{A'}{\bra{(m,\bar{n}))}}.
    \end{split}
\end{equation}

\indent Note that $\Pi_A \Pi_{A \cup B} = \Pi_{A \cup B} \Pi_A = \Pi_{A \cup B}$, and similarly for $\Pi_B$, because of frustration-freeness. Frustration-freeness also implies that $\Pi_A \wedge \Pi_B = \Pi_{A \cup B}$. \newline
\indent Now, let's consider the quantity of interest
\begin{equation}\label{eq: definition of delta}
    \begin{split}
        \delta(A, B) \coloneq \| \Pi_A \Pi_B - \Pi_A \wedge \Pi_B \| = \| \Pi_A \Pi_B - \Pi_{A \cup B} \| = \| (\Pi_A - \Pi_{A \cup B})(\Pi_B - \Pi_{A \cup B})\|.
    \end{split}
\end{equation}
In order to prove this bound, we consider an equivalent characterization of $\delta(A,B)$ in terms of states instead of projections:
\begin{equation}\label{eq: wavefunction definition of delta}
\begin{split}
    \delta(A,B) = \mathrm{sup}\{ |\braket{\psi}{\chi}|: &\ket{\psi}, \ket{\chi} \in \mathcal{H}_{A \cup B}, \braket{\psi}{\psi} \leq 1, \braket{\chi}{\chi} \leq 1, \\
    &(\Pi_A - \Pi_{A \cup B}) \ket{\psi} = \ket{\psi}, (\Pi_B - \Pi_{A \cup B})\ket{\chi} = \ket{\chi}\}.
\end{split}
\end{equation}
Note that the condition $(\Pi_A - \Pi_{A \cup B}) \ket{\psi} = \ket{\psi}$ means that $\ket{\psi}$ is a ground state in $A$ and it is orthogonal to the ground states in $A \cup B$, and similarly for $\ket{\chi}$. So we can write them as
\begin{equation}
    \begin{split}
        &\ket{\psi} = \ket{+}_A \ket{\psi_{+}}_{B'} + \sum_{m,n = \pm 1} \ket{(m,n)}_A \ket{\psi_{(m,n)}}_{B'},\\
        &\ket{\chi} = \ket{\chi_+}_{A'} \ket{+}_{B} + \sum_{m,n = \pm1} \ket{\chi_{(m,n)}}_{A'}\ket{(m,n)}_B,
    \end{split}
\end{equation}
for some states $\ket{\psi_+}, \ket{\psi_{(m,n)}} \in \mathcal{H}_{B'}$ and $\ket{\chi_+}, \ket{\chi_{(m,n)}} \in \mathcal{H}_{A'}$.\newline
\indent Restricting $\ket{\psi}$ and $\ket{\chi}$ to be orthogonal to the ground states in $A \cup B$ gives the following relations
\begin{equation}\label{eq: orthongonal condition for psi}
    \begin{split}
        0 &= \prescript{}{A \cup B} {\braket{+}{\psi}} = \prescript{}{B'}{\braket{+}{\psi_{+}}_{B'}} + \frac{1}{2^{|A| / 2}} \sum_{m,n = \pm 1} (-1)^{mn} \prescript{}{B'}{\braket{+}{\psi_{(m,n)}}_{B'}} \\
        0 &= \prescript{}{A \cup B} {\braket{(m,n)}{\psi}} = \prescript{}{A \cup B}{\braket{(m,n)}{+}_A} \ket{\psi_+}_{B'} + \sum_{m',n' = \pm 1} \prescript{}{A \cup B}{\braket{(m,n)}{(m',n')}_A} \ket{\psi_{(m',n')}}_{B'} \\
         &= \frac{1}{2^{|A|/2}} \prescript{}{B'}{\braket{(m,n)}{\psi_{+}}_{B'}} + \sum_{m',n' = \pm 1} \prescript{}{B'}{\bra{(1,1)}} \prescript{}{A}{\bra{(m,n)}} CZ_{\beta - 1, \beta} \ket{(m,n)}_A\ket{\psi_{(m',n')}}_{B'} \\
         &= \frac{1}{2^{|A|/2}} \prescript{}{B'}{\braket{(m,n)}{\psi_{+}}_{B'}} + \sum_{m',n' = \pm 1} \prescript{}{B'}{\bra{(1,1)}} (\frac{1}{2}\delta_{m,m'}\delta_{n,n'} + \frac{1}{2}Z_{\beta} \delta_{m,m'}\delta_{n,n'} + \frac{1}{2}\delta_{m,m'}\delta_{n,\bar{n}'} - \frac{1}{2}Z_{\beta} \delta_{m,m'}\delta_{n,\bar{n}'}) \ket{\psi_{(m',n')}}_{B'}\\
         &= \frac{1}{2^{|A|/2}} \prescript{}{B'}{\braket{(m,n)}{\psi_{+}}_{B'}} + \frac{1}{2} \prescript{}{B'}{\bra{(1, 1)} \ket{\psi_{(m,n)}}_{B'}} + \frac{1}{2} \prescript{}{B'}{\braket{(-1, 1)}{\psi_{(m,n)}}_{B'}} + \frac{1}{2} \prescript{}{B'}{\bra{(1, 1)} \ket{\psi_{(m,\bar{n})}}_{B'}} - \frac{1}{2} \prescript{}{B'}{\bra{(-1, 1)} \ket{\psi_{(m,\bar{n})}}_{B'}}
    \end{split}
\end{equation}
Similarly, we have
\begin{equation}\label{eq: orthongonal condition for chi}
    \begin{split}
        0 &= \prescript{}{A \cup B}{\braket{+}{\chi}} = \prescript{}{A'}{\bra{+}\ket{\psi_+}_{A'}} + \frac{1}{2^{|B|/2}}\sum_{m,n = \pm 1} (-1)^{mn} \prescript{}{A'}{\bra{+}} \ket{\chi_{(m,n)}}_{A'} \\
        0 &= \prescript{}{A\cup B} {\braket{+}{\chi}} \\
        &= \frac{1}{2^{|B|/2}} \prescript{}{A'}{\braket{(m,n)}{\chi_{+}}_{A'}} + \frac{1}{2} \prescript{}{A'}{\bra{(1, 1)}} \ket{\chi_{(m,n)}}_{A'} + \frac{1}{2} \prescript{}{A'}{\braket{(1, -1)}{\chi_{(m,n)}}_{A'}} + \frac{1}{2} \prescript{}{A'}{\bra{(1, 1)} \ket{\chi_{(\bar{m},n)}}_{A'}} - \frac{1}{2} \prescript{}{A'}{\bra{(1, -1)} \ket{\chi_{(\bar{m},n)}}_{A'}}
    \end{split}
\end{equation}
The constraints on the norms give
\begin{equation} \label{eq: norm constraints}
    \begin{split}
        1 &\geq \braket{\psi}{\psi} \\
        &= \prescript{}{B'}{\braket{\psi_+}{\psi_+}_{B'}} + \sum_{m,n} \prescript{}{B'}{\braket{\psi_{(m,n)}}{\psi_{(m,n)}}_{B'}} + \frac{1}{2^{|A|/2}} \sum_{m,n} \left( \prescript{}{B'}{\braket{\psi_+}{\psi_{(m,n)}}_{B'}} + \prescript{}{B'}{\braket{\psi_{(m,n)}}{\psi_{+}}_{B'}} \right) \\
        &\geq (1 - \frac{4}{2^{|A|/2}}) \prescript{}{B'}{\braket{\psi_+}{\psi_+}_{B'}} + (1 - \frac{1}{2^{|A|/2}})\sum_{m,n} \prescript{}{B'}{\braket{\psi_{(m,n)}}{\psi_{(m,n)}}_{B'}} \\
        \\
        1 &\geq \braket{\chi}{\chi} \\
        &= \prescript{}{A'}{\braket{\chi_+}{\chi_+}_{A'}} + \sum_{m,n} \prescript{}{A'}{\braket{\chi_{(m,n)}}{\chi_{(m,n)}}_{A'}} + \frac{1}{2^{|B|/2}} \sum_{m,n} \left( \prescript{}{A'}{\braket{\chi_+}{\chi_{(m,n)}}_{A'}} + \prescript{}{A'}{\braket{\chi_{(m,n)}}{\chi_{+}}_{A'}} \right) \\
        &\geq (1 - \frac{4}{2^{|B|/2}}) \prescript{}{A'}{\braket{\chi_+}{\chi_+}_{A'}} + (1 - \frac{1}{2^{|B|/2}})\sum_{m,n} \prescript{}{A'}{\braket{\chi_{(m,n)}}{\chi_{(m,n)}}_{A'}}.
    \end{split}
\end{equation}
The overlap between $\ket{\psi}$ and $\ket{\chi}$ is
\begin{equation}
    \begin{split}
        \braket{\psi}{\chi} &= \prescript{}{A'}{\braket{+}{\chi_{+}}_{A'}} \prescript{}{B'}{\braket{\psi_{+}}{+}_{B'}} + \sum_{m,n} \prescript{}{A}{\braket{(m,n)}{+}_{A \cap B}}\ket{\chi_{+}}_{A'} \prescript{}{B'}{\braket{\psi_{(m,n)}}{+}_{B'}} \\
        &+ \sum_{m,n} \prescript{}{A'}{\braket{+}{\chi_{(m,n)}}_{A'}} \prescript{}{B'}{\bra{\psi_{+}}} \prescript{}{A \cap B}{\braket{+}{(m,n)}_{B}} + \sum_{m,n, m', n'} \prescript{}{B'}{\bra{\psi_{m',n'}}} \prescript{}{A}{\braket{(m',n')}{(m,n)}_{B}} \ket{\chi_{(m,n)}}_{A'} \\
        &= \frac{1}{2^{(\abs{A} + \abs{B})/2}} \sum_{m,n,m',n'} (-1)^{mn + m'n'}\prescript{}{A'}{\braket{+}{\chi_{(m,n)}}_{A'}} \prescript{}{B'}{\braket{\psi_{(m',n')}}{+}_{B'}} \\
        &+ \frac{1}{2^{\abs{A \cap B}/ 2}} \sum_{m,n} \prescript{}{A'}{\braket{(m,n)}{\chi_{+}}_{A'}} \prescript{}{B'}{\braket{\psi_{(m,n)}}{+}_{B'}} \\
        &+ \frac{1}{2^{\abs{A \cap B}/ 2}} \sum_{m,n} \prescript{}{A'}{\braket{+}{\chi_{m}}_{A'}} \prescript{}{B'}{\braket{\psi_{+}}{(m,n)}_{B'}} \\
        &+ \sum_{m,n, m', n'} \prescript{}{B'}{\bra{\psi_{m',n'}}} \prescript{}{A}{\braket{(m',n')}{(m,n)}_{B}} \ket{\chi_{(m,n)}}_{A'},
    \end{split}
\end{equation}
where we used \eqref{eq: orthongonal condition for psi} and \eqref{eq: orthongonal condition for chi} to obtain the first line of the second equation. Now our goal is to pull out an exponentially small factor from the fourth line of the second equation, then we can bound $\abs{\braket{\psi}{\chi}}$ by a sum of exponentially small numbers.\newline
\indent We first compute $\prescript{}{A}{\braket{(m',n')}{(m,n)}_{B}}$ in the fourth line and it can be written as
\begin{equation}
    \begin{split}
        \prescript{}{A}{\braket{(m',n')}{(m,n)}_{B}} &= \prescript{}{A'}{}{\bra{(1,1)}} \prescript{}{A \cap B}{\bra{(m',n')}} CZ_{\alpha, \alpha + 1} CZ_{\beta - 1, \beta} \ket{(m,n)}_{A \cap B} \ket{(1,1)}_{B'} \\
        &=\prescript{}{A'}{\bra{(1,1)}} \Big( \frac{1}{4}\delta_{m,m'}\delta_{n,n'}(1 + Z_{\alpha} + Z_{\beta} + Z_{\alpha} Z_{\beta}) + \frac{1}{4}\delta_{\bar{m}',m}\delta_{n',n}(1 - Z_{\alpha} + Z_{\beta} - Z_{\alpha} Z_{\beta}) \\
        &+ \delta_{m',m}\delta_{\bar{n}',n}(1 + Z_{\alpha} - Z_{\beta} - Z_{\alpha} Z_{\beta}) + \frac{1}{4}\delta_{\bar{m}',m}\delta_{\bar{n}',n}(1 - Z_{\alpha} - Z_{\beta} + Z_{\alpha} Z_{\beta}) \Big) \ket{(1,1)}_{B'}
    \end{split}
\end{equation}
Then the fourth line becomes
\begin{equation}
    \begin{split}
        \sum_{m,n, m', n'} &\prescript{}{B'}{\bra{\psi_{m',n'}}} \prescript{}{A}{\braket{(m',n')} {(m,n)}_{B}} \ket{\chi_{(m,n)}}_{A'} \\
        = \frac{1}{4} \sum_{m,n} &\textcolor{ForestGreen}{\prescript{}{A'}{\braket{(1,1)}{\chi_{(m,n)}}_{A'}} \prescript{}{B'}{\braket{\psi_{(m,n)}}{(1,1)}_{B'}}} + \textcolor{red}{\prescript{}{A'}{\braket{(1,-1)}{\chi_{(m,n)}}_{A'}} \prescript{}{B'}{\braket{\psi_{(m,n)}}{(1,1)}_{B'}}}\\
        &+ \textcolor{ForestGreen}{\prescript{}{A'}{\braket{(1,1)}{\chi_{(m,n)}}_{A'}} \prescript{}{B'}{\braket{\psi_{(m,n)}}{(-1,1)}_{B'}}} + \textcolor{red}{\prescript{}{A'}{\braket{(1,-1)}{\chi_{(m,n)}}_{A'}} \prescript{}{B'}{\braket{\psi_{(m,n)}}{(1,-1)}_{B'}}} \\
        &+ \textcolor{orange}{\prescript{}{A'}{\braket{(1,1)}{\chi_{(m,n)}}_{A'}} \prescript{}{B'}{\braket{\psi_{(\bar{m}, n)}}{(1,1)}_{B'}}} -\textcolor{violet}{\prescript{}{A'}{\braket{(1,-1)}{\chi_{(m,n)}}_{A'}} \prescript{}{B'}{\braket{\psi_{(\bar{m}, n)}}{(1,1)}_{B'}}} \\
        &+ \textcolor{orange}{\prescript{}{A'}{\braket{(1,1)}{\chi_{(m,n)}}_{A'}} \prescript{}{B'}{\braket{\psi_{(\bar{m}, n)}}{(-1,1)}_{B'}}} - \textcolor{violet}{\prescript{}{A'}{\braket{(1,-1)}{\chi_{(m,n)}}_{A'}} \prescript{}{B'}{\braket{\psi_{(\bar{m}, n)}}{(-1,1)}_{B'}}} \\
        &+ \textcolor{ForestGreen}{\prescript{}{A'}{\braket{(1,1)}{\chi_{(m,n)}}_{A'}} \prescript{}{B'}{\braket{\psi_{(m,\bar{n})}}{(1,1)}_{B'}}} + \textcolor{red}{\prescript{}{A'}{\braket{(1,-1)}{\chi_{(m,n)}}_{A'}} \prescript{}{B'}{\braket{\psi_{(m,\bar{n})}}{(1,1)}_{B'}}} \\
        &- \textcolor{ForestGreen}{\prescript{}{A'}{\braket{(1,1)}{\chi_{(m,n)}}_{A'}} \prescript{}{B'}{\braket{\psi_{(m,\bar{n})}}{(1,-1)}_{B'}}} - \textcolor{red}{\prescript{}{A'}{\braket{(1,-1)}{\chi_{(m,n)}}_{A'}} \prescript{}{B'}{\braket{\psi_{(m,\bar{n})}}{(-1,1)}_{B'}}} \\
        &+ \textcolor{orange}{\prescript{}{A'}{\braket{(1,1)}{\chi_{(m,n)}}_{A'}} \prescript{}{B'}{\braket{\psi_{(\bar{m},\bar{n})}}{(1,1)}_{B'}}} - \textcolor{violet}{\prescript{}{A'}{\braket{(1,-1)}{\chi_{(m,n)}}_{A'}} \prescript{}{B'}{\braket{\psi_{(\bar{m},\bar{n})}}{(1,1)}_{B'}}} \\
        &- \textcolor{orange}{\prescript{}{A'}{\braket{(1,1)}{\chi_{(m,n)}}_{A'}} \prescript{}{B'}{\braket{\psi_{(\bar{m},\bar{n})}}{(-1,1)}_{B'}}} + \textcolor{violet}{\prescript{}{A'}{\braket{(1,-1)}{\chi_{(m,n)}}_{A'}} \prescript{}{B'}{\braket{\psi_{(\bar{m},\bar{n})}}{(-1,1)}_{B'}}} \\
        = \frac{1}{4} \sum_{m,n} &\textcolor{ForestGreen}{\prescript{}{A'}{\braket{(1,1)}{\chi_{(m,n)}}_{A'}} \Big( \prescript{}{B'}{\braket{\psi_{(m,n)}}{(1,1)}_{B'}} + \prescript{}{B'}{\braket{\psi_{(m,n)}}{(-1,1)}_{B'}} + \prescript{}{B'}{\braket{\psi_{(m,\bar{n})}}{(1,1)}_{B'}} - \prescript{}{B'}{\braket{\psi_{(m,\bar{n})}}{(1,-1)}_{B'}} \Big)} \\
        &+ \textcolor{red}{\prescript{}{A'}{\braket{(1,-1)}{\chi_{(m,n)}}_{A'}}\Big(\prescript{}{B'}{\braket{\psi_{(m,n)}}{(1,1)}_{B'}} + \prescript{}{B'}{\braket{\psi_{(m,n)}}{(1,-1)}_{B'}} + \prescript{}{B'}{\braket{\psi_{(m,\bar{n})}}{(1,1)}_{B'}} - \prescript{}{B'}{\braket{\psi_{(m,\bar{n})}}{(-1,1)}_{B'}} \Big)} \\
        &+ \textcolor{orange}{\prescript{}{A'}{\braket{(1,1)}{\chi_{(m,n)}}_{A'}}\Big( \prescript{}{B'}{\braket{\psi_{(\bar{m}, n)}}{(1,1)}_{B'}} + \prescript{}{B'}{\braket{\psi_{(\bar{m}, n)}}{(-1,1)}_{B'}} + \prescript{}{B'}{\braket{\psi_{(\bar{m},\bar{n})}}{(1,1)}_{B'}} - \prescript{}{B'}{\braket{\psi_{(m,n)}}{(-1,1)}_{B'}} \Big)} \\
        &- \textcolor{violet}{\prescript{}{A'}{\braket{(1,-1)}{\chi_{(m,n)}}_{A'}}\Big( \prescript{}{B'}{\braket{\psi_{(\bar{m}, n)}}{(1,1)}_{B'}} + \prescript{}{B'}{\braket{\psi_{(\bar{m}, n)}}{(-1,1)}_{B'}} + \prescript{}{B'}{\braket{\psi_{(\bar{m},\bar{n})}}{(1,1)}_{B'}} - \prescript{}{B'}{\braket{\psi_{(\bar{m},\bar{n})}}{(-1,1)}_{B'}} \Big)},
    \end{split}
\end{equation}
where we colored the terms that will be grouped together, and the summation inside each parenthesis can be simplified using \eqref{eq: orthongonal condition for psi}. Then, we get
\begin{equation}
    \begin{split}
        \sum_{m,n, m', n'} &\prescript{}{B'}{\bra{\psi_{m',n'}}} \prescript{}{A}{\braket{(m',n')} {(m,n)}_{B}} \ket{\chi_{(m,n)}}_{A'} \\
        =\frac{1}{2} \frac{1}{2^{\abs{A}/2}} \sum_{m,n} &-\prescript{}{A'}{\braket{(1,1)}{\chi_{(m,n)}}_{A'}} \prescript{}{B'}{\braket{\psi_+}{(m,n)}_{B'}} - \prescript{}{A'}{\braket{(1,-1)}{\chi_{(m,n)}}} \prescript{}{B'}{\braket{\psi_+}{m, n}} \\
        &- \prescript{}{A'}{\braket{(1,1)}{\chi_{(m,n)}}_{A'}}  \prescript{}{B'}{\braket{\psi_+}{(\bar{m},n)}_{B'}} + \prescript{}{A'}{\braket{(1,-1)}{\chi_{(m,n)}}_{A'}}\prescript{}{B'}{\braket{\psi_+}{(\bar{m},n)}_{B'}} \\
        = \frac{1}{2} \frac{1}{2^{\abs{A}/2}} \sum_{m,n} &\Big(-\prescript{}{A'}{\braket{(1,1)}{\chi_{(m,n)}}_{A'}} - \prescript{}{A'}{\braket{(1,-1)}{\chi_{(m,n)}}}- \prescript{}{A'}{\braket{(1,1)}{\chi_{(\bar{m},n)}}_{A'}} + \prescript{}{A'}{\braket{(1,-1)}{\chi_{(\bar{m},n)}}_{A'}}\Big) \prescript{}{B'}{\braket{\psi_+}{(m,n)}_{B'}} \\
        = \frac{1}{2^{(\abs{A} + \abs{B}) / 2}} \sum_{m,n} & \prescript{}{A'} {\braket{(m,n)}{\chi_+}_{A'}} \prescript{}{B'}{\braket{\psi_+}{(m,n)}_{B'}} .
    \end{split}
\end{equation}
Finally the overlap between $\ket{\psi}$ and $\ket{\chi}$ is given by
\begin{equation}
    \begin{split}
        \braket{\psi}{\chi} &= \frac{1}{2^{(\abs{A} + \abs{B})/2}} \sum_{m,n,m',n'} (-1)^{mn + m'n'} \prescript{}{A'}{\braket{+}{\chi_{(m,n)}}_{A'}} \prescript{}{B'}{\braket{\psi_{(m',n')}}{+}_{B'}} \\
        &+ \frac{1}{2^{\abs{A \cap B}/ 2}} \sum_{m,n} \prescript{}{A'}{\braket{(m,n)}{\chi_{+}}_{A'}} \prescript{}{B'}{\braket{\psi_{(m,n)}}{+}_{B'}} \\
        &+ \frac{1}{2^{\abs{A \cap B}/ 2}} \sum_{m,n} \prescript{}{A'}{\braket{+}{\chi_{m}}_{A'}} \prescript{}{B'}{\braket{\psi_{+}}{(m,n)}_{B'}} \\
        &+ \frac{1}{2^{(\abs{A} + \abs{B}) / 2}} \sum_{m,n} \prescript{}{A'}{\braket{(m,n)}{\chi_+}_{A'}} \prescript{}{B'}{\braket{\psi_+}{(m,n)}_{B'}}.\\
    \end{split}
\end{equation}
The absolute value of $\braket{\psi}{\chi}$ can thus be bounded by
\begin{equation}
    \begin{split}
        \abs{\braket{\psi}{\chi}} &\leq \frac{1}{2^{(\abs{A} + \abs{B})/2}} \sum_{m,n,m',n'} \abs{\prescript{}{A'}{\braket{+}{\chi_{(m,n)}}_{A'}}}\abs{\prescript{}{B'}{\braket{\psi_{(m',n')}}{+}_{B'}}} \\
        &+ \frac{1}{2^{\abs{A \cap B}/ 2}} \sum_{m,n} \abs{\prescript{}{A'}{\braket{(m,n)}{\chi_{+}}_{A'}}} \abs{ \prescript{}{B'}{\braket{\psi_{(m,n)}}{+}_{B'}}} \\
        &+ \frac{1}{2^{\abs{A \cap B}/ 2}} \sum_{m,n} \abs{\prescript{}{A'}{\braket{+}{\chi_{m}}_{A'}}} \abs{\prescript{}{B'}{\braket{\psi_{+}}{(m,n)}_{B'}}} \\
        &+ \frac{1}{2^{(\abs{A} + \abs{B}) / 2}} \sum_{m,n} \abs{\prescript{}{A'}{\braket{(m,n)}{\chi_+}_{A'}}} \abs{\prescript{}{B'}{\braket{\psi_+}{(m,n)}_{B'}}}.
    \end{split}
\end{equation}
We then try to bound each line separately. For the first line, we have
\begin{equation}
    \begin{split}
        &\sum_{m,n,m',n'} \abs{\prescript{}{A'}{\braket{+}{\chi_{(m,n)}}_{A'}}}\abs{\prescript{}{B'}{\braket{\psi_{(m',n')}}{+}_{B'}}} \\
        &\leq \sqrt{2\sum_{m', n'} \abs{\prescript{}{A'}{\braket{+}{\chi_{(m,n)}}_{A'}}}^2} \sqrt{2 \sum_{m,n}\abs{\prescript{}{B'}{\braket{\psi_{(m',n')}}{+}_{B'}}}^2 } \\
        &\leq 2 \sqrt{\sum_{m',n'} \prescript{}{A'}{\braket{\chi_{(m',n')}}{\chi_{(m',n')}}_{A'}}}  \sqrt{ \sum_{m,n}\abs{\prescript{}{B'}{\braket{\psi_{(m',n')}}{\psi_{(m',n')}}_{B'}}} } \\
        &\leq \frac{2}{\sqrt{(1 - \frac{1}{2^{\abs{B} / 2}})(1 - \frac{1}{2^{\abs{A}/2}})}} \\
        &\leq 4,
    \end{split}
\end{equation}
where we used the norm constraints \eqref{eq: norm constraints} to get the third inequality, and assumed that $A$ and $B$ are large enough to get the last inequality. The second line can be bounded as
\begin{equation}
    \begin{split}
        &\sum_{m,n} \prescript{}{A'}{\braket{(m,n)}{\chi_{+}}_{A'}} \prescript{}{B'}{\braket{\psi_{(m,n)}}{+}_{B'}} \\
        &\leq \sqrt{\sum_{m,n} \abs{\prescript{}{A'}{\braket{(m,n)}{\chi_{+}}_{A'}}}^2} \sqrt{\sum_{m,n}\abs{\prescript{}{B'}{\braket{\psi_{(m,n)}}{+}_{B'}}}^2} \\
        &\leq \sqrt{\prescript{}{A'}{\braket{\chi_{+}}{\chi_{+}}_{A'}}} \sqrt{\sum_{m,n}\prescript{}{B'}{\braket{\psi_{(m,n)}}{\psi_{(m,n)}}_{B'}}}\\
        &\leq \frac{1}{\sqrt{(1 - \frac{4}{2^{|B| / 2}})(1 - \frac{1}{2^{|A|/2}})}} \\
        &\leq 2.
    \end{split}
\end{equation}
Similarly, we have
\begin{equation}
    \sum_{m,n} \abs{\prescript{}{A'}{\braket{+}{\chi_{m}}_{A'}}} \abs{\prescript{}{B'}{\braket{\psi_{+}}{(m,n)}_{B'}}} \leq 2.
\end{equation}
Finally, the last line reads
\begin{equation}
    \begin{split}
        &\sum_{m,n} \abs{\prescript{}{A'}{\braket{(m,n)}{\chi_+}_{A'}}} \abs{\prescript{}{B'}{\braket{\psi_+}{(m,n)}_{B'}}} \\
        &\leq \sum_{m,n,m',n'} \abs{\prescript{}{A'}{\braket{(m,n)}{\chi_+}_{A'}}} \abs{\prescript{}{B'}{\braket{\psi_+}{(m',n')}_{B'}}} \\
        &\leq \sqrt{2 \sum_{m,n} \abs{\prescript{}{A'}{\braket{(m,n)}{\chi_+}_{A'}}}^2} \sqrt{2 \sum_{m',n'} \abs{\prescript{}{B'}{\braket{\psi_+}{(m',n')}_{B'}}}^2} \\
        &\leq 2 \sqrt{\prescript{}{A'}{\braket{\chi_+}{\chi_+}_{A'}}} \sqrt{\prescript{}{B'}{\braket{\psi_+}{\psi_+}_{B'}}} \\
        &\leq \frac{2}{\sqrt{(1 - \frac{4}{2^{|B|/2}})(1 - \frac{4}{2^{|A|/2}})}} \\
        &\leq 4.
    \end{split}
\end{equation}
Combining those inequalities, we have
\begin{equation}
    \abs{\braket{\psi}{\chi}} \leq \frac{8}{2^{(\abs{A} + \abs{B}) / 2}} + \frac{4}{2^{|A \cap B|/2}},
\end{equation}
for all sufficiently large $A$ and $B$. Given the definition of $\delta(A,B)$ in \eqref{eq: wavefunction definition of delta}, it follows that
\begin{equation}
    \delta(A,B) \leq \frac{8}{2^{(\abs{A} + \abs{B}) / 2}} + \frac{4}{2^{|A \cap B|/2}}.
\end{equation}

\subsection{Proof of local indistinguishability}\label{sec:local indistinguishability}

Here we prove that the two states $ \frac{1}{\sqrt{Z}_{\pm} } \left( \ket{+} \pm 
 \ket{\text{cluster}}\right)$ are locally indistinguishable in the thermodynamic limit.

First, we fix the normalization constant $Z_\pm $: 
\begin{equation}
\begin{split}
Z_\pm &= \left(    \bra{+} \pm 
 \bra{\text{cluster}} \right)  \left( \ket{+} \pm 
 \ket{\text{cluster}} \right) \\
&=2 \pm  \bra{+} \ket{\text{cluster}} \pm \bra{\text{cluster}}  \ket{+}
\end{split}
\end{equation}
With $\bra{+}\ket{\text{cluster}} =\bra{\text{cluster}} \ket{+} = 2^{-\frac{N}{2}+1  }$,  we find $Z_{\pm} = 2 \pm  2^{-\frac{N}{2} +2   }$, which becomes $Z_\pm =2$ in the thermodynamic limit $N \to \infty$.

We proceed to compute the reduced density matrix on an interval $A$ by tracing out its complement $\overline{A}$: 

\begin{equation}
\rho_{A,\pm} =\frac{1}{Z_{\pm}} \ket{+} \bra{+}_{A} + \frac{1}{Z_{\pm}}  \tr_{\overline{A}} \left( \ket{\text{cluster}  }  \bra{ \text{cluster} }\right) \pm \frac{1}{Z_{\pm}} \tr_{\overline{A}}  \left( \ket{+} \bra{\text{cluster}}\right)  \pm  \frac{1}{Z_{\pm}} \left(\tr_{\overline{A}} \ket{ 
 \text{cluster}} \bra{+}\right) 
\end{equation}

For the third term: 
\begin{equation}
\tr_{\overline{A}}  \left( \ket{+} \bra{\text{cluster}}\right) = \tr_{\overline{A}}  \left(  \ket{+}\bra{+}_A\otimes  \ket{+}\bra{+}_{\overline{A}}U_{CZ,A}U_{CZ,\overline{A}} U_{CZ,\partial}       \right), 
\end{equation}
where $U_{CZ,A}, U_{CZ,\overline{A}}, U_{CZ,\partial} $ denotes the product of controlled-Z acting only in $A, \overline{A}$, and the bipartition boundary. To proceed:

\begin{equation}
\begin{split}
\tr_{\overline{A}}  \left( \ket{+} \bra{\text{cluster}}\right) &=  \ket{+}\bra{+}_A U_{CZ,A}\tr_{\overline{A}}  \left(  \ket{+}\bra{+}_{\overline{A}}U_{CZ,\overline{A}} U_{CZ,\partial}       \right) \\
&=  \ket{+}\bra{+}_A U_{CZ,A}  \bra{+}_{\overline{A}} U_{CZ,\overline{A}} U_{CZ,\partial}   \ket{+}_{\overline{A}}, 
\end{split}
\end{equation}
and we will show that this quantity decays exponentially with the size of $\overline{A}$. 

To start, we label the qubits in the interval $\overline{A}$ by the index $i=1,\cdots ,n$, where $n$ is the total number of qubits in $\overline{A}$. Then $\bra{+}_{\overline{A}} U_{CZ,\overline{A}} U_{CZ,\partial}   \ket{+}_{\overline{A}}$ can be computed in the computational basis ($\sigma_i = 0, 1$): 

\begin{equation}
\bra{+}_{\overline{A}} U_{CZ,\overline{A}} U_{CZ,\partial}   \ket{+}_{\overline{A}} = 2^{-n} \sum_{ \sigma_1, \cdots, \sigma_N}  (-1)^{\sigma_0 \sigma_1+ \sigma_1\sigma_2 + \cdots + \sigma_N\sigma_{N+1}},    
\end{equation}
where $\sigma_0, \sigma_{N+1}  $ denotes the two 
qubits (in the computational basis) in $A$ on the two end-points of $\overline{A}$. There are two cases corresponding to even $n$ or odd $n$.

\noindent (1)~ Even n: in this case, when summing over the odd spins in $\overline{A}$, i.e.  $\sigma_1, \sigma_3, \cdots, \sigma_{n-1}$, we have 

\begin{equation}
\bra{+}_{\overline{A}} U_{CZ,\overline{A}} U_{CZ,\partial}   \ket{+}_{\overline{A}} = 2^{-\frac{N}{2} }  \sum_{\sigma_2, \sigma_4 \cdots, \sigma_N} \delta( \sigma_0=\sigma_2 = \sigma_4 =\cdots = \sigma_N   ) (-1)^{\sigma_N \sigma_{N+1}} =  2^{-\frac{N}{2} } (-1)^{\sigma_0 \sigma_{N+1}}
\end{equation}
In terms of operators, one finds 
\begin{equation}
\bra{+}_{\overline{A}} U_{CZ,\overline{A}} U_{CZ,\partial}   \ket{+}_{\overline{A}} = 2^{-\frac{N}{2} }   CZ_{0,N+1}
\end{equation}
which decays exponentially in the size of the $\overline{A}$.

\noindent (2)~ Odd n: in this case, when summing over the odd spins in $\overline{A}$, i.e.  $\sigma_1, \sigma_3, \cdots, \sigma_{n-1}$, we have 

\begin{equation}
\bra{+}_{\overline{A}} U_{CZ,\overline{A}} U_{CZ,\partial}   \ket{+}_{\overline{A}} = 2^{-\frac{n+1}{2} }  \sum_{\sigma_2, \sigma_4 \cdots, \sigma_{n-1}} \delta( \sigma_0=\sigma_2 = \sigma_4 =\cdots = \sigma_{n+1}   ) =  2^{-\frac{n+1}{2} }  \delta(\sigma_0= \sigma_{n+1}). 
\end{equation}
In terms of operators, one finds 
\begin{equation}
\bra{+}_{\overline{A}} U_{CZ,\overline{A}} U_{CZ,\partial}   \ket{+}_{\overline{A}} = 2^{-\frac{n+1}{2} }  \frac{1+Z_0Z_{n+1}  }{2}
\end{equation}
which decays exponentially in the size of the $\overline{A}$.

In either case 1 or 2, the quantity decays exponentially, and therefore, one finds the local indistinguishability when the subregion $A$ is of constant size.

\begin{equation}
\boxed{
\norm{\rho_{A,+} - \rho_{A,-}}_1 \leq  O(e^{-\alpha N}) }
\end{equation}
with $\alpha$ being a $O(1)$ constant.

\subsection{Calculation of order parameters}\label{appendix:1d_order}
Define the local operator $ O_i =X_i (1- Z_{i-1} Z_{i+1} )$ charged under $U_{\text{CZ}}$, we compute the two-point function w.r.t. $\ket{\psi} = \frac{1}{\sqrt{2}} ( \ket{ + }^{\otimes N }  + \ket{\text{cluster}} )  $:

\begin{equation}
\begin{split}
\expval{O_i O_j } =& \frac{1}{2} [
\bra{+}^{\otimes N  } O_i O_j \ket{+}^{\otimes N}+  \bra{\text{cluster}} O_i O_j  \ket{\text{cluster}} \\
&+ \bra{+}^{\otimes N  } O_i O_j   \ket{\text{cluster}} + \bra{\text{cluster}} O_i O_j \ket{+}^{\otimes N} ]
\end{split}
\end{equation}

We have $\bra{+}^{\otimes N  } O_i O_j \ket{+}^{\otimes N}= \bra{\text{trivial}}  (1-Z_{i-1}Z_{i+1}) (1-Z_{j-1}Z_{j+1})   \ket{+}^{\otimes N}= 1$ for $i \neq j$. Similarly, $\bra{\text{cluster}} O_i O_j  \ket{\text{cluster}} = 1$ for $i \neq j$. In the thermodynamic limit, the last two terms vanish, and thus $\expval{O_iO_j}=1$ for any $i \neq j$.  
\subsection{Calculation of disorder parameters}\label{appendix:1d_disorder}
Here we consider the disorder operator $\prod_{i=1}^{n-1} \text{CZ}_{i,i+1}$ and calculate its expectation value w.r.t. $\ket{\psi}=\frac{1}{\sqrt{Z}}  \left( \ket{+}^{\otimes N}  + \ket{\text{cluster}}  \right)$:
\begin{equation}
\begin{split}
\expval{\prod_{i=1}^{n-1} \text{CZ}_{i,i+1} } =& \frac{1}{Z} [
\bra{+}^{\otimes N  } \prod_{i=1}^{n-1} \text{CZ}_{i,i+1} \ket{+}^{\otimes N}+  \bra{\text{cluster}} \prod_{i=1}^{n-1} \text{CZ}_{i,i+1}  \ket{\text{cluster}} \\
&+ \bra{+}^{\otimes N  } \prod_{i=1}^{n-1} \text{CZ}_{i,i+1}   \ket{\text{cluster}} + \bra{\text{cluster}} \prod_{i=1}^{n-1} \text{CZ}_{i,i+1} \ket{+}^{\otimes N} ] \\=
&\frac{2}{Z} \left[
\bra{+}^{\otimes N  } \prod_{i=1}^{n-1} \text{CZ}_{i,i+1} \ket{+}^{\otimes N}    + \bra{+}^{\otimes N  } \prod_{i=1}^{n-1} \text{CZ}_{i,i+1}   \ket{\text{cluster}} \right]
\end{split}
\end{equation}

\noindent \underline{Calculation of $\bra{+}^{\otimes N  } \prod_{i=1}^{n-1} \text{CZ}_{i,i+1} \ket{+}^{\otimes N}  $:}

This quantity can be calculated as
\begin{equation}
\bra{+}^{\otimes n } \prod_{i=1}^{n-1} \text{CZ}_{i,i+1} \ket{+}^{\otimes n}=  2^{-n} \sum_{\{\sigma_i | i=1,2,3,\cdots, n\}}  (-1)^{\sum_{i=1}^{n-1} \sigma_i \sigma_{i+1}  }
\end{equation}
with each $\sigma_i =0,1$. 

\begin{itemize}
    \item Even $n$: \\
Summing over $\sigma_i$ with $i=1,3,5,\cdots, n-1$ gives

\begin{equation}
\bra{+}^{\otimes n } \prod_{i=1}^{n-1} \text{CZ}_{i,i+1} \ket{+}^{\otimes n}= 2^{-n}2^{\frac{n}{2}} \sum_{\{\sigma_i | i=2,4,6,\cdots, n\}} \delta(0= \sigma_2=\sigma_4=\cdots= \sigma_n) =2^{-\frac{n}{2}}. 
\end{equation}

\item Odd $n$: \\
Summing over $\sigma_i$ with $i=1,3,5,\cdots, n$ gives

\begin{equation}
\bra{+}^{\otimes n } \prod_{i=1}^{n-1} \text{CZ}_{i,i+1} \ket{+}^{\otimes n}= 2^{-n}2^{\frac{n+1}{2}} \sum_{\{\sigma_i | i=2,4,6,\cdots, n-1\}} \delta(0= \sigma_2=\sigma_4=\cdots= \sigma_n-1) =2^{\frac{-n+1}{2}}. 
\end{equation}
\end{itemize}

\noindent\underline{Calculation of $\bra{+}^{\otimes N  } \prod_{i=1}^{n-1} \text{CZ}_{i,i+1} \ket{\text{cluster}}$:}\\

Since $\ket{\text{cluster}} =\prod_{i=1}^{N} \text{CZ}_{i,i+1}  \ket{+}^{\otimes N  }$, one has

\begin{equation}
\begin{split}
\bra{+}^{\otimes N  } \prod_{i=1}^{n-1} \text{CZ}_{i,i+1} \ket{\text{cluster}}&=    \bra{+}^{\otimes N  } \prod_{i=n}^{N}    \text{CZ}_{i,i+1} \ket{+}^{\otimes N  }\\
&= \bra{+}^{\otimes N-n+2  } \prod_{i=n}^{N}    \text{CZ}_{i,i+1} \ket{+}^{\otimes N-n+2} \\ 
&= \begin{cases}
2^{-\frac{N-n}{2} -1} \quad \text{for even n}\\
2^{ -\frac{N-n}{2} -\frac{1}{2}  }   \quad \text{for odd n}
\end{cases}
\end{split}
\end{equation}

Combining the results above, one finds 

\begin{equation}
\expval{\prod_{i=1}^{n-1} \text{CZ}_{i,i+1} } =
\begin{cases}
\frac{2}{Z}  \left[ 2^{-\frac{n}{2}} + 2^{-\frac{N-n}{2} -1}     \right] \quad ~~~\text{for even n}\\
\frac{2}{Z}  \left[ 2^{\frac{-n+1}{2}} +    2^{ -\frac{N-n}{2} -\frac{1}{2}  }   \right] \quad \text{for odd n}
\end{cases}
\end{equation}
Therefore, if we first take the thermodynamic limit ($N\to \infty$), the expectation value of the disorder becomes
\begin{equation}
\expval{\prod_{i=1}^{n-1} \text{CZ}_{i,i+1} } =
\begin{cases}
2^{-\frac{n}{2}}  \quad ~~~\text{for even n}\\
2^{\frac{-n+1}{2}} \quad \text{for odd n}
\end{cases}
\end{equation}
which decays exponentially in the size of the disorder operator.

\subsection{Numerics on the parent Hamiltonian}\label{appendix:parent_1d}
Here we provide additional numerical results for the Hamiltonian 

\begin{equation}
\begin{split}
H'= -2 \sum_i (X_i + Z_{i-1}X_{i} Z_{i+1}) +(1+g)\sum_i(Z_{i}X_{i+1}Z_{i+2}X_{i+3} +X_{i}Z_{i+1}X_{i+2}Z_{i+3}) 
    \end{split}
\end{equation}

\begin{figure}[h!]
\includegraphics[width=\textwidth]{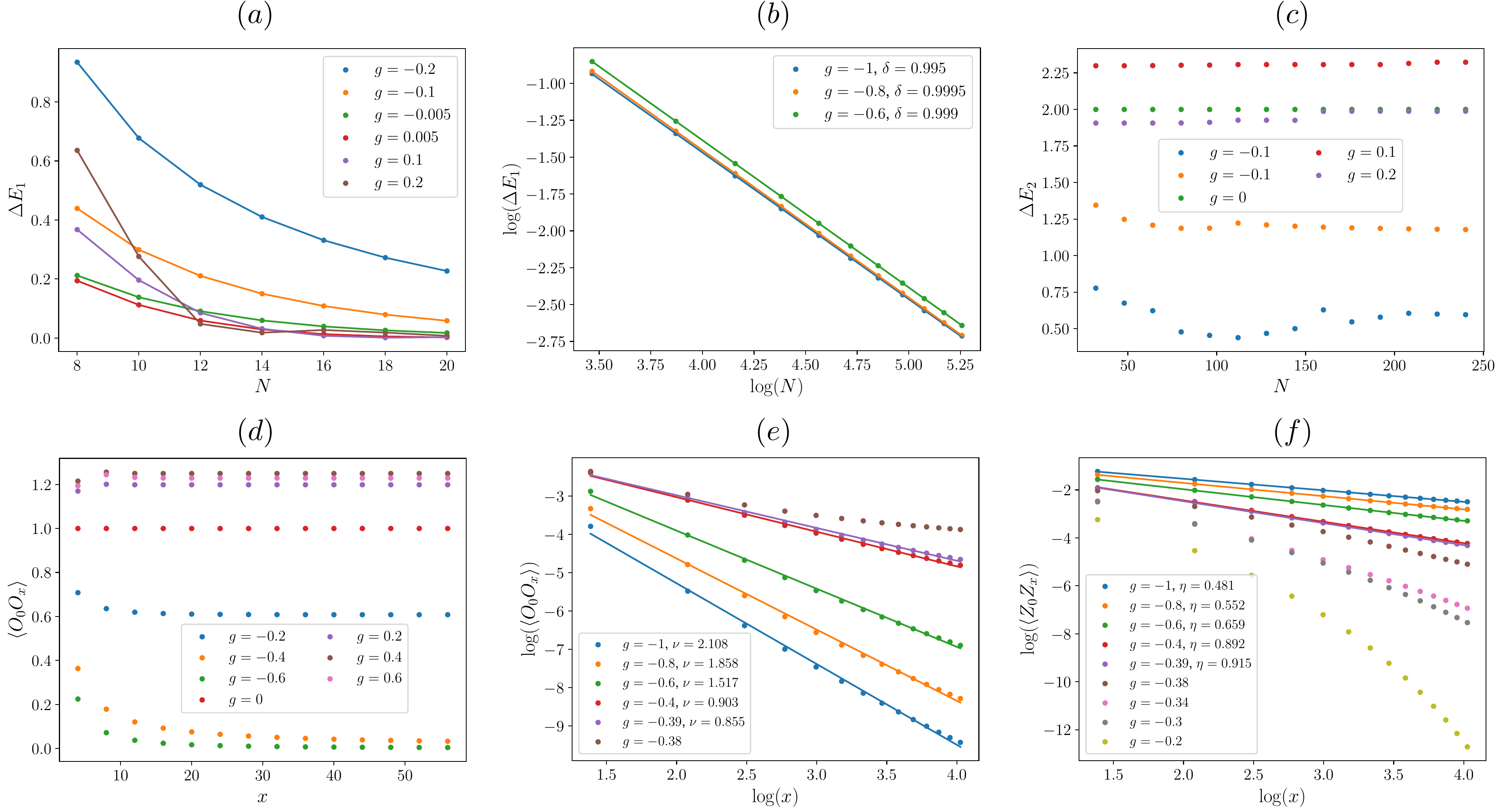}
    \caption{(a) Exact Diagonalization results for the energy gap $\Delta E_1$, the energy difference of the two lowest eigenstates, versus total system size $N$ in the non-onsite symmetry breaking phase. $\Delta E_1$ decreases with $N$, presumably will decay to zero and give two-fold ground state degeneracy in the thermodynamic limit. (b) Log-log plot of $\Delta E_1$ versus $N$ at various $g$ in the CFT phase. By fitting the data with the function $\frac{1}{N^{\delta}}$, we find $\delta$ very close to $1$, consistent with the gap scaling of the CFT phase. (c) $\Delta E_2$, the energy difference between the second lowest and the third lowest eigenstates, versus $N$ for various $g$ in the non-onsite symmetry breaking phase. The data suggest a constant energy gap when increasing $N$. (d) Correlation function $\expval{O_0 O_x}$ of the non-invertible charged operators versus distance $x$ at a fixed $N = 240$ for various $g$. In the non-onsite symmetry-breaking phase ($g \gtrapprox -0.4$), the correlation approaches a constant as increasing $N$, which implies a long-range order. In the CFT phase ($g\lessapprox -0.4$), the correlation decays to zero. (e) In the log-log plot of $\expval{O_0 O_x}$ in the CFT phase, one finds the algebraic correlation $\expval{O_0 O_x} \sim x^{- \nu}$. (f) Log-log plot of $\expval{Z_0Z_x}$ versus distance $x$ at $N = 240$ for various $g$. In the CFT phase, $\expval{Z_0 Z_x}$ exhibits an algebraic decay $\sim x^{-\eta}$. In the non-onsite symmetry breaking phase, $\expval{Z_0 Z_x}$ decays exponentially to $0$.}
    \label{fig:numerics}    
\end{figure}

\subsection{Stability of $c=1$ CFT from bosonization}\label{append:boson}
In this section, we discuss the field theory describing the long-wavelength limit of the Hamiltonian in Eq.\eqref{eq:1d_hamiltonian_appendix} at $g = -1$, which takes the form
\begin{equation}
    H_0 = \sum_{i} -Z_{i-1} X_{i} Z_{i + 1} - X_{i}.
\end{equation}
As has been pointed out in \cite{Levin_gu_spt_2012, tantivasadakarn2023pivot, pace2025lattice}, $H_0$ at long-wavelength is described by a compact boson CFT at radius $r= 1 /2$ (or equivalently Luttinger parameter $K = 1/4$). If we denote the compact boson field as $\phi(x)$ and its dual field as $\theta(x)$, the theory has a $U(1)$ momentum symmetry generated by $\partial_x \theta$. One can identify the $U(1)$ momentum symmetry in the IR to a $U(1)$ symmetry of $H_0$ whose generator is given by $G_{U(1)} = \frac{1}{4}\sum_{i} (-1)^i Z_{i}Z_{i+1}$. Notice that since $e^{-i \pi G_{U(1)}} = \prod_{i} \mathrm{CZ}_{i,i+1}$, so the non-onsite CZ symmetry generates the transformation $\phi \rightarrow \phi + \pi, \theta \rightarrow \theta$ in the IR. The other two onsite symmetries $U_0 = \prod_i X_{2i+1}$ and $U_e = \prod_i X_{2i}$ generate the transformation $\phi \xrightarrow{} -\phi, \theta \xrightarrow{} -\theta$ and $\phi \xrightarrow{} -\phi, \theta \xrightarrow{} \pi -\theta$ respectively in the IR.

However, as we perturb away from $g = -1$, $G_{U(1)}$ no longer commutes with $H$ but $e^{-i \pi G_{U(1)}}$ remains a symmetry of $H$. Here we present a field-theoretic argument about the stability of the $c=1$ CFT phase around $g = -1$, which is the same as saying there can still be a $U(1)$ symmetry emergent in the IR. First, when the model is slightly perturbed away from $g= - 1$, the Hamiltonian takes the form: 
\begin{equation}
    \begin{split}
\sum_i - Z_{i}X_{i+1}Z_{i+2} - X_{i+1} - \lambda Z_{i}X_{i+1}Z_{i+2}X_{i+3} - \lambda X_{i}Z_{i+1}X_{i+2}Z_{i+3},
    \end{split}
\end{equation}
where $\lambda$ controls the perturbation strength. Using a KT transformation \cite{Kennedy_Tasaki_1992,2023_KT_oshikawa}, defined by the following mapping of local operators: 

\begin{equation}
\begin{split}
Z_{i-1} X_iZ_{i+1 } &\to  Z_{i-1} Z_{i+1 }  \\
X_i  &\to X_i \\
\end{split}
\end{equation}
The above Hamiltonian can be mapped to 
\begin{equation}
    \begin{split}
        H(\lambda)
        &= \sum_i - Z_{2i}Z_{2i+2} - X_{2i} - Z_{2i+1}Z_{2i+3} - X_{2i+1} + \lambda Z_{2i}Z_{2i+2} X_{2i+3} + \lambda X_{2i} Z_{2i+1} Z_{2i+3} \\
        &+ \lambda Z_{2i+1}Z_{2i+3}X_{2i+4} + \lambda X_{2i+1}Z_{2i+2}Z_{2i+4},
    \end{split}
\end{equation}
which is described by two decoupled critical Ising chains (living on even and odd sublattices) perturbed by three-body interactions. The two decoupled Ising CFTs can be bosonized into a compact boson \cite{giamarchi2003quantum}, where we can identify the following mapping from Ising local operators to compact boson operators (to the leading order)
\begin{equation}
\begin{split}
    &Z_{2i} Z_{2i + 1} \sim \sigma(x,\tau) \tilde{\sigma}(x,\tau) \sim \cos(\phi(x,\tau)) \\
    &X_{2i} X_{2i+1} \sim Z_{2i} Z_{2i+1} Z_{2i+2} Z_{2i+3} \sim \mathcal{E}(x,\tau) + \tilde{\mathcal{E}}(x,\tau) \sim \cos(2\phi(x, \tau)),
\end{split}
\end{equation}
where the spin operator $\sigma(x,\tau)$ has scaling dimension $\frac{1}{8}$ and the energy density operator $\mathcal{E}$ has scaling dimension $1$ in Ising CFT.\newline
\indent If we further apply a Krammers-Wannier transformation to the odd sublattice, we obtain the following Hamiltonian
\begin{equation}
\begin{split}
    H(\lambda)
        &= \sum_i - Z_{2i}Z_{2i+2} - X_{2i} - Z_{2i+1}Z_{2i+3} - X_{2i+1} + \lambda Z_{2i}Z_{2i+2} Z_{2i+3} Z_{2i+5} + \lambda X_{2i} X_{2i+3} \\
        &+ \lambda X_{2i+3}X_{2i+4} + \lambda Z_{2i+1}Z_{2i+2}Z_{2i+3}Z_{2i+4}
    \end{split}
\end{equation}
The perturbations $X_{2i+3}X_{2i+4} + Z_{2i+1}Z_{2i+2}Z_{2i+3}Z_{2i+4}$ generates a marginal perturbation $\cos(2\phi(x, \tau))$ that varies the boson compactification radius continuously \cite{fabrizio2000critical}. 
The perturbations $X_{2i} X_{2i+3} + Z_{2i}Z_{2i+2} Z_{2i+3} Z_{2i+5}$, which differ the formers by only a lattice translation on some operators, would at most generate the marginal perturbations $\cos(2\phi(x,\tau))$ and the rest would be of higher-order irrelevant terms. By this perturbative argument on the scaling dimensions of the perturbation, we therefore expect a finite range of the $c=1$ CFT phase around $g = -1$, which is also consistent with our numerical calculation of the central charge. A detailed renormalization group study of this marginal perturbation is left for future work. \newline

\noindent \textbf{Remarks on the critical exponents}: here we present some additional remarks on the critical exponents of the correlation: $\expval{Z_0 Z_x}$ and $\expval{O_0 O_x}$ (for even $x$). At $g=-1$, using the KT transformation, these two correlations can be mapped to $\expval{Z_0 X_1 X_3...X_{x-1} Z_{x}}$ and $\expval{\left(X_0 - Z_{N-1}Z_1 \right)\left(X_x - Z_{x-1}Z_{x+1} \right)}$ with respect to the ground state of two decoupled critical Ising chains on the two sublattices. We then apply the following lattice to continuum correspondence for the primary fields
\begin{equation}
    \begin{split}
        Z_x \sim \sigma(x),~~~Z_{x} Z_{x+2} \sim 1 + \mathcal{E}(x),~~~ X_x \sim 1 - \mathcal{E}(x),  ~~~X_0 X_2 X_4...X_x \sim \mu(x),
    \end{split}
\end{equation}
where the same mapping also applies to lattice operators on the odd sites and we use $\widetilde{\sigma}$, $\widetilde{\mu}$, and $\widetilde{\mathcal{E}}$ to denote their corresponding Ising primary fields. Then, the two correlation functions can be simplified to correlation functions of Ising primary fields
\begin{equation}
    \begin{split}
        &\expval{Z_0 X_1 X_3...X_{x-1} Z_{x}} = \expval{Z_0 Z_x}\expval{X_1 X_3...X_{x-1}} \sim \expval{\sigma(0) \sigma(x)} \expval{\widetilde{\mu}(0) \widetilde{\mu}(x)}\sim \frac{1}{x^{1/2}} \\
        &\expval{\left(X_0 - Z_{N-1}Z_1 \right)\left(X_x - Z_{x-1}Z_{x+1} \right)} \sim \expval{\mathcal{E}(0)\mathcal{E}(x)} + \expval{\widetilde{\mathcal{E}}(0)\widetilde{\mathcal{E}}(x)}\sim \frac{1}{x^2}.
    \end{split}
\end{equation}
The above two critical exponents $\frac{1}{2}, 2$ are close to the exponents $0.481$ and $2.108$ obtained from our numerics, see Fig.\ref{fig:numerics}(e)(f). As we turn on the perturbation $\lambda$, numerically we see the critical components are continuously varying. This agrees with our previous argument that the perturbation is exactly marginal and would modify the compactification radius of the $c=1$ CFT.

We also want to conjecture that the transition at $g \approx $ 0.4 between the critical phase and the non-onsite symmetry-breaking phase is the $SU(2)_1$ CFT, which is the simplest CFT with a $Z_2^3$ type anomaly. For such a theory, the only non-trivial primary operator has a scaling dimension $\frac{1}{2}$, which is consistent with our numerical result that both the $\left\langle Z_0 Z_x \right\rangle$ and $\left\langle O_0 O_x\right\rangle$ correlations have a critical exponent $\nu \approx 1$. We left the detailed analysis of this critical point to future work.

\subsection{State preparation via measurement and feedback}\label{appendix:preparation}

Here we provide a measurement-feedback protocol (i.e. adaptive circuit) to prepare the state $\ket{+}^{\otimes N } + 
\ket{\text{cluster}}$ with probability  $\frac{1}{2}$ in the thermodynamic limit. 

\begin{figure}[h!]
    \centering
    \includegraphics[width=0.8\linewidth]{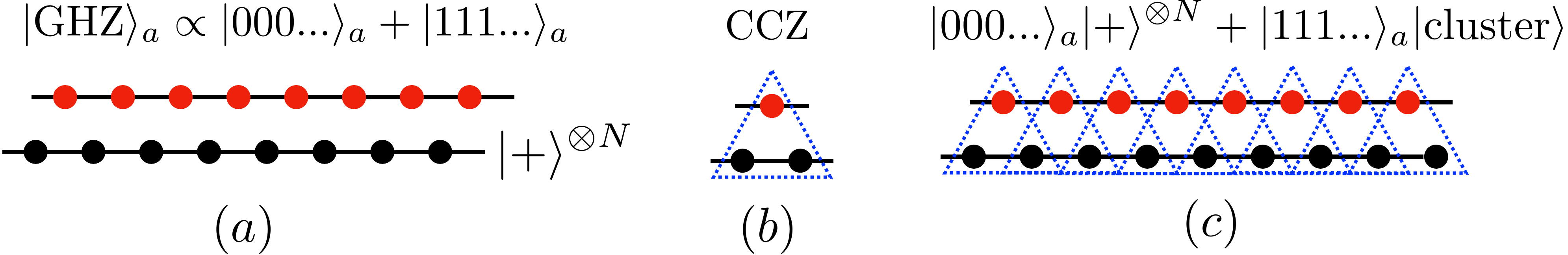}
    \caption{Measurement-based state-preparation protocol:  
(a) We first add a chain of ancilla qubits in the GHZ state, which can be prepared in constant depth with measurement and feedback. (b) A single CCZ gate ($| 0\rangle_a \langle 0|_a +  | 1\rangle_a \langle 1 |_a \text{CZ}_{12}$) acts on two physical qubits and one ancilla qubit in between. (c) After extensively applying CCZ gates, performing single-site X measurements on all ancilla qubits produces $\ket{+}^{\otimes N } + \ket{\text{cluster}}$ with probability $\frac{1}{2}$ in the thermodynamic limit.} 
\end{figure}

To start, we initialize the $N$ physical qubits in the $\ket{+}$ state. On each bond connecting two physical qubits, we place an ancilla qubit in $\ket{+}$ and transform these $N$ ancilla qubits into a GHZ state via a constant-depth adaptive circuit (the generation of the GHZ state can achieved by measuring a cluster state followed by unitary feedback \cite{Raussendorf_2001_ghz}, or just measuring all the neighboring two-qubit $ZZ$ Pauli operators followed by unitary feedback). It follows that the physical and ancilla qubits are in the state $\ket{000...}_a   \ket{+}^{\otimes N}  +\ket{111...}_a  \ket{+}^{\otimes N}$. Now we apply a layer of unitary gates, each of which is a CCZ gate acting on the two neighboring physical qubits and one ancilla qubit in between. This gives state $\ket{000...}_a\ket{+}^{\otimes N}  +\ket{111...}_a \prod_{i=1}^{N} \text{CZ}_{i,i+1} \ket{+}^{\otimes N}=\ket{000...}_a\ket{+}^{\otimes N}  +\ket{111...}_a \ket{\text{cluster}}$. 

Now we perform the single-qubit $X$ measurement on all the ancilla qubits. Denoting the measurement outcome by $\{x_{i+\frac{1}{2}}\}$, a corresponding pure-state trajectory is $\ket{+}^{\otimes N}  +  \prod_{i=1}^{N} x_{i+\frac{1}{2}  }  \ket{\text{cluster}}$. As such, the sign is solely given by global parity of the measurement outcome $\prod_{i=1}^{N} x_{i+\frac{1}{2}}$, and in the thermodynamic limit, the probability having $\prod_{i=1}^{N} x_{i+\frac{1}{2}} = 1, -1 $ is $\frac{1}{2}$. As a result, the state ($\ket{+}^{\otimes N} +  \ket{\text{cluster}}$) can be prepared in constant depth with a constant probability $\frac{1}{2}$ in the thermodynamic limit. Below we make two remarks: \\

\noindent (1) While the above protocol produces $ \ket{\psi_{\pm}} \propto \ket{+} \pm  \ket{\text{cluster}}$ where the sign $\pm$ determined from the measurement outcome is probabilistic, the reduced matrix on a local region is, in fact, deterministically prepared since $\ket{\psi_{\pm}}$ is locally indistinguishable in thermodynamic limit. Therefore, post-selection is not needed for experimentally measuring the local order operators.\\

\noindent (2) The protocol can be generalized to prepare the superposition of any $O(1)$ number of short-range entangled pure states (i.e. the states that can be prepared by constant-depth local unitary circuits) with $O(1)$ success probability. For instance, we can prepare the superposition of a $\mathbb{Z}_2$ symmetric product state (i.e. $\ket{+}^{\otimes N}$) and the $\mathbb{Z}_2$ Levin-Gu SPT state \cite{Levin_gu_spt_2012}. Such a state can further be deterministically converted to a superposition of the ($\mathbb{Z}_2$ topologically ordered) toric code state and the (twisted $\mathbb{Z}_2$ topologically ordered) double semion state by a Kramer-Wannier duality, which can be implemented by measurement and feedback in constant depth \cite{ashvin_2021_measurement}.

\section{2d SPT soup}

\subsection{Derivation of the SPT soup wavefunction} \label{sec: SPT soup derivation}
Here we show that implementing measurements (with post-selecting the measurement outcomes) on a $\mathbb{Z}_{2}^3$ SPT gives rise to the SPT soup wavefunction defined in the main text.

Following Ref.\cite{yoshida_2016_symmetry}, we consider a triangular lattice, with each vertex accommodating a qubit. The fixed-point $\mathbb{Z}_{2}^3$ SPT is given by $\ket{\psi_{Z_{2}^3}} \equiv \prod_{\triangle} \mathrm{CCZ}_{\triangle} \ket{+}$, where $\mathrm{CCZ}$ denotes the controlled-controlled-Z gate and the product is taken over all small triangles on the lattice. Since the triangular lattice is 3-colorable, one can define the sublattices $A, B, C$ so that all neighboring qubits belong to different sublattices. It follows that $\ket{\psi_{Z_{2}^3}}$ respects the following three global (0-form) $\mathbb{Z}_2$ symmetries acting on the three sublattices:

\begin{equation}
    U_A = \prod_{i\in A } X_{i} ~~~U_B = \prod_{i\in B} X_{i} ~~~U_C = \prod_{i\in C} X_{C}.
\end{equation}

The structure of the $Z_{2}^3$ SPT wavefunction can be understood from the decorated domain wall picture\cite{chen_decorated}, which we review below. Consider a small patch on the triangular lattice involving four sites: 
\begin{equation}
\includegraphics[height=6em]{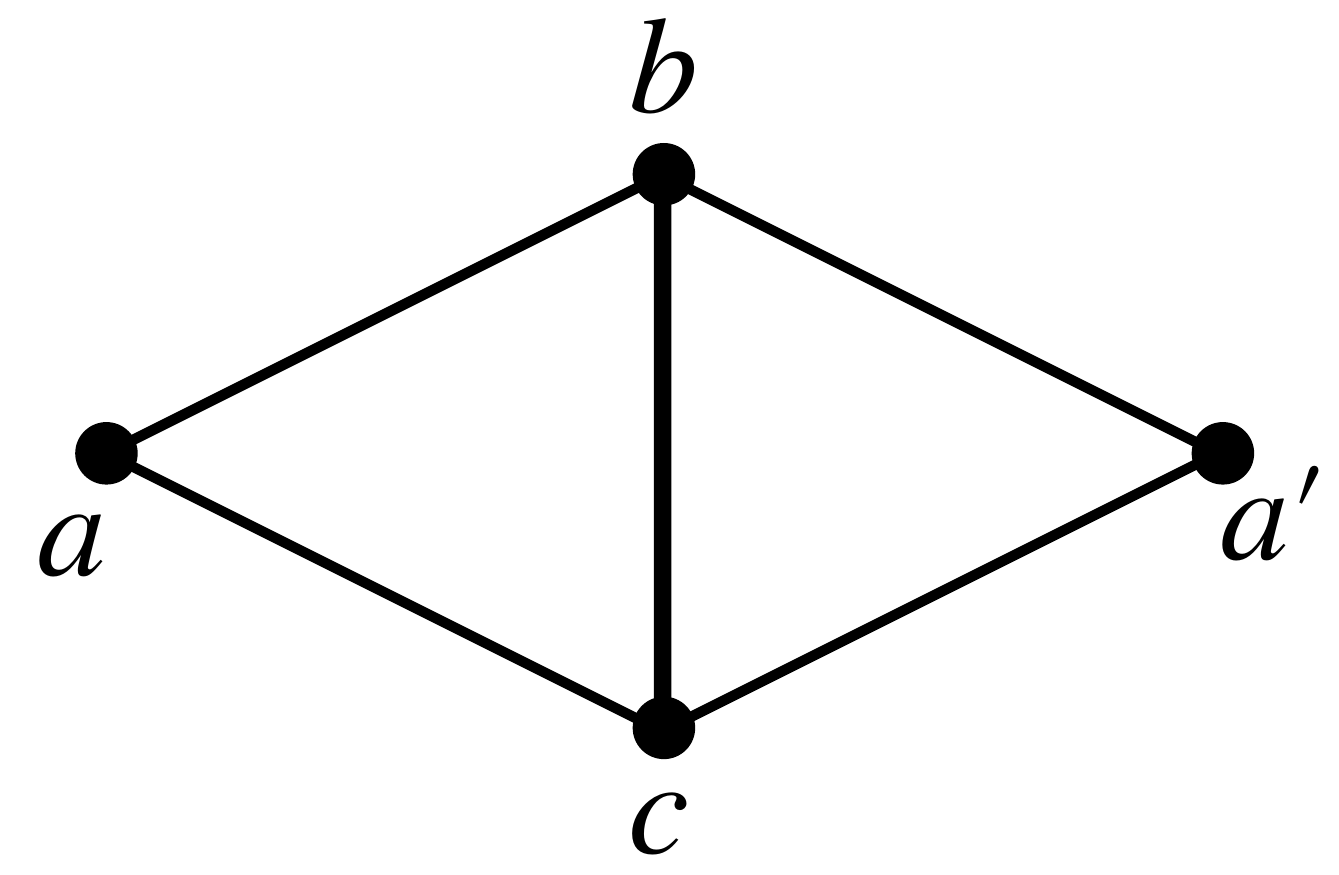}
\end{equation}
where the sites $a$ and $a'$ belong to the $A$ sublattice; $b$ and $c$ belong to $B$ and $C$ sublattices respectively. All four sites are initiated in the $\ket{+}$ state and there are CCZ gates applied to $(a, b, c)$ and $(a', b, c)$. The effect of the CCZ gates on $b$ and $c$ can be seen by expanding $a$ and $a'$ in the $Z$ basis
\begin{equation}
    \begin{split}
        \mathrm{CCZ}_{(abc)} \mathrm{CCZ}_{(a'bc)} &\ket{Z_a = s} \ket{Z_{a'} = s'} = \left[  \delta_{s, s'} + (1 - \delta_{s, s'})\mathrm{CZ}_{(bc)}\right] \ket{Z_a = s} \ket{Z_{a'} = s'}.
    \end{split}
\end{equation}
This shows that if there is a domain wall between $a$ and $a'$ (i.e. $s\neq s'$), then one applies a CZ gate on that domain wall, which is the edge connecting $b$ and $c$. With this four-qubit example, one can now understand the structure of the $Z_{2}^3$ SPT wavefunction and write it as 

\begin{equation}
    \ket{\psi_{Z_{2}^3}} \propto \sum_{s} \ket{s}_A \prod_{\expval{ i \in B, j \in C } \in \mathcal{C}}\text{CZ}_{ij }  \ket{+}_{BC}, 
\end{equation}
where $\mathcal{C}$ denotes the loops (on the BC edges) that correspond to the domain-wall configurations of $s$. Namely, the $Z_{2}^3$ SPT can be understood as a condensate of A domain walls, where each domain wall is decorated by a 1d cluster state living on the sublattices $B$ and $C$. 

Based on this structure, one can immediately see that by measuring the Pauli $Z$ on all qubits on the $A$ sublattice with an outcome $s$, which also fixes the domain-wall pattern $C(s)$, the post-measurement state on $B, C$ reads 

\begin{equation}\label{eq:Z_overlap}
\begin{split}
_A\bra{s}  \ket{\psi_{Z_{2}^3}} \propto \prod_{\expval{ i \in B, j \in C } \in \mathcal{C}(s)}\text{CZ}_{ij }  \ket{+}_{BC} = \ket{\text{SPT}}_{\mathcal{C}(s)} \ket{+}_{\overline{\mathcal{C}}(s)}. 
\end{split}
\end{equation}
where $\ket{\text{SPT}}_{\mathcal{C}(s)}$ is the 1d cluster-state SPT defined on $B, C$ along the domain wall $\mathcal{C}(s)$, and all the other qubits not on the domain wall, i.e. $\overline{\mathcal{C}}(s)$, are in the Pauli-X product state $\ket{+}$.

This allows us to easily derive the state obtained by Pauli-X measurements on the sublattice $A$. Specifically, the single-qubit Pauli-X measurement on the sublattice $A$ (with the fixed measurement outcomes $X_{i}=1$) projects all the qubits on $A$ into $\ket{+}_A$, and the corresponding post-measurement pure-state on $B, C$ is:

\begin{equation}
~_A\bra{+}  \ket{\psi_{Z_{2}^3}} \propto \sum_{s} ~_A\bra{s}  \ket{\psi_{Z_{2}^3}}  \propto \sum_{\mathcal{C}} \ket{\mathrm{SPT}}_{\mathcal{C}} \ket{+}_{\overline{\mathcal{C}}}  
\end{equation}
where we first expand $_A\bra{+}$ in the Z basis, and then use Eq.\ref{eq:Z_overlap}. This is exactly the SPT soup discussed in the main text.

\subsection{Levin-Gu SPT under measurements}\label{sec:levin_gu}
With a similar strategy, one can also derive the post-measurement wavefunction obtained by measuring Pauli-X on one sublattice for the Levin-Gu SPT \cite{Levin_gu_spt_2012} (also known as the $Z_2$ SPT).   On a triangular lattice, the Levin-Gu SPT is defined as $\ket{\psi_{Z_2}} = U_{\mathrm{CCZ}} U_{\mathrm{CZ}} U_{\mathrm{Z}} \ket{+}$ \cite{liu2022many}, where ${\mathrm{CCZ}}$ denotes CCZ gates applied to every triangle, $U_{\mathrm{CZ}}$ denotes CZ gates applied to every edge, and $U_{\mathrm{Z}}$ denotes Pauli-Zs applied to every site. Again, we first divide the triangular lattice into the sublattices $A, B, C$, and perform Pauli-X measurements on every qubit on the sublattice $A$. The post-measurement state on $B, C$ reads
\begin{equation}
\ket{\phi} \propto  _A\bra{+} \ket{\psi_{Z_2}} \propto \sum_{s} \prescript{}{A}{\bra{s'}} \prod_{\triangle} \mathrm{CCZ}_{\triangle} \prod_{\left\langle ij\right\rangle} \mathrm{CZ}_{ij} \prod_{i} Z_i \ket{+}
\end{equation}

To proceed, we first note that
\begin{equation}
_A\bra{s}  \prod_{\triangle} \mathrm{CCZ}_{\triangle}  =~ _A\bra{s}  \prod_{\expval{ij} \in \mathcal{C}(s)} \mathrm{CZ}_{ij},
\end{equation}
where   $\mathcal{C}(s)$ denotes the domain wall of the configuration $s$ on $A$ sublattice. On the other hand, 
\begin{equation}
\begin{split}
_A\bra{s} \prod_{\expval{ij}} \mathrm{CZ}_{ij}& =    \left[\prod_{\expval{ij} \in B \cup C}   \mathrm{CZ}_{ij}\right]   ~_A\bra{s} \prod_{\expval{i \in A, j \in B }}  \mathrm{CZ}_{ij}   \prod_{\expval{i \in A, j \in C }}  \mathrm{CZ}_{ij} \\
&=   \left[\prod_{\expval{ij} \in B \cup C}   \mathrm{CZ}_{ij}\right]   \prod_{i \in B \cup C} Z_{i}^{\frac{1 - \prod_{j \in \text{nn}(i), j \in A} s_j}{2}}  ~_A\bra{s},
\end{split}
\end{equation} 
where $\prod_{j \in \text{nn}(i), j \in A} s_j$ denotes the product of three $s$ on the three sites (belonging to $A$) that are the nearest neighbors of the site $i$ (belonging to either $B$ or $C$). Combining the above results and using $_A\bra{s} \prod_{i\in A } Z_i =  _A\bra{s} \prod_{i\in A } s_i $, one finds

\begin{equation}
\ket{\phi} \propto   \left[\prod_{\expval{ij} \in B \cup C}   \mathrm{CZ}_{ij}\right] \left[  \prod_{i  \in B \cup C}Z_i  \right] \sum_{s}  \prod_{\expval{ij} \in \mathcal{C}(s)}  \mathrm{CZ}_{ij} \prod_{i \in B \cup C} Z_{i}^{\frac{1 - \prod_{j \in \text{nn}(i), j \in A} s_j}{2}}  \prod_{i\in A } s_i \ket{+}_{BC}
\end{equation}

This can be further simplified by noticing that $s$ and $\overline{s} \equiv \{-s_i, \forall i \}$ have the same domain wall configuration $\mathcal{C}$, i.e. $\mathcal{C}(s) = \mathcal{C}(\overline{s})$. This allows us to rewrite the summation over $s$ as the summation over $\mathcal{C}$:
\begin{equation}\label{eq:sum_C}
    \begin{split}
&\sum_{s}  \prod_{\expval{ij} \in \mathcal{C}(s)}  \mathrm{CZ}_{ij} \prod_{i \in B \cup C} Z_{i}^{\frac{1 - \prod_{j \in \text{nn}(i), j \in A} s_j}{2}}  \prod_{i\in A } s_i  \\
&= \sum_{\mathcal{C}} \left[   \prod_{\expval{ij} \in \mathcal{C}(s)}  \mathrm{CZ}_{ij} \prod_{i \in B \cup C} Z_{i}^{\frac{1 - \prod_{j \in \text{nn}(i), j \in A} s_j}{2}}  \prod_{i\in A } s_i  +      \prod_{\expval{ij} \in \mathcal{C}(s)}  \mathrm{CZ}_{ij} \prod_{i \in B \cup C} Z_{i}^{\frac{1 - \prod_{j \in \text{nn}(i), j \in A} \overline{s}_j}{2}}  \prod_{i\in A } \overline{s}_i      \right]
\end{split}
\end{equation}

We consider the case with an even number of sites on the sublattice $A$, so $\prod_{i\in A } s_i = \prod_{i\in A } \overline{s}_i$. Also, $Z_{i}^{\frac{1 -\prod_{j \in \text{nn}(i), j \in A} \overline{s}_j}{2}} = Z_i ~Z_i^{\frac{1 - \prod_{j \in \text{nn}(i), j \in A} s_j}{2}}$. Therefore, Eq.\ref{eq:sum_C} can be simplified as 

\begin{equation}
 \left(1 + \prod_{i \in B \cup C} Z_{i}\right) \sum_{\mathcal{C}}   \prod_{\expval{ij} \in \mathcal{C}(s)}  \mathrm{CZ}_{ij} \prod_{i \in B \cup C} Z_{i}^{\frac{1 - \prod_{j \in \text{nn}(i), j \in A} s_j}{2}}  \prod_{i\in A } s_i.  
 \end{equation}

Finally, the post-measurement state can be written as 

\begin{equation}
\ket{\phi}\propto   \left[\prod_{\expval{ij} \in B \cup C}   \mathrm{CZ}_{ij}\right]  \left(1 + \prod_{i \in B \cup C} Z_{i}\right) \sum_{\mathcal{C}}   \prod_{\expval{ij} \in \mathcal{C}(s')}  \mathrm{CZ}_{ij} \prod_{i \in B \cup C} Z_{i}^{\frac{1 - \prod_{j \in \text{nn}(i), j \in A} s_j}{2}}  \prod_{i\in A } s_i  \ket{+}_{BC}
\end{equation}

This can be further simplified as 
\begin{equation}
    \begin{split}
\ket{\phi} \propto    \left[\prod_{\expval{ij} \in B \cup C}   \mathrm{CZ}_{ij}\right]  \left(1 + \prod_{i \in B \cup C} Z_{i}\right) \prod_{p} \left( 1 + \widetilde{U}_p \right) \ket{+}_{B  C},
    \end{split}
\end{equation}
where $\widetilde{U}_p$ is an operator acting on the 6 qubits on the vertices of a plaquette $p$ on the $B \cup C$: 
\begin{equation}
\widetilde{U}_p \equiv \prod_{\expval{ij} \in p} \mathrm{CZ}_{ij} \prod_{i \in p} \left[(-1)^{\frac{1}{6}}Z_{i}\right]
\end{equation}

This shows that $\ket{\phi}$ respects the 1-form non-onsite symmetry generated by $\widetilde{U}_p $, namely $\widetilde{U}_p \ket{\phi} = \ket{\phi}$. Taking the product of $\widetilde{U}_p$ generates the 1-form symmetry along any contractible loops, e.g.

\begin{equation*}
\includegraphics[width=7cm]{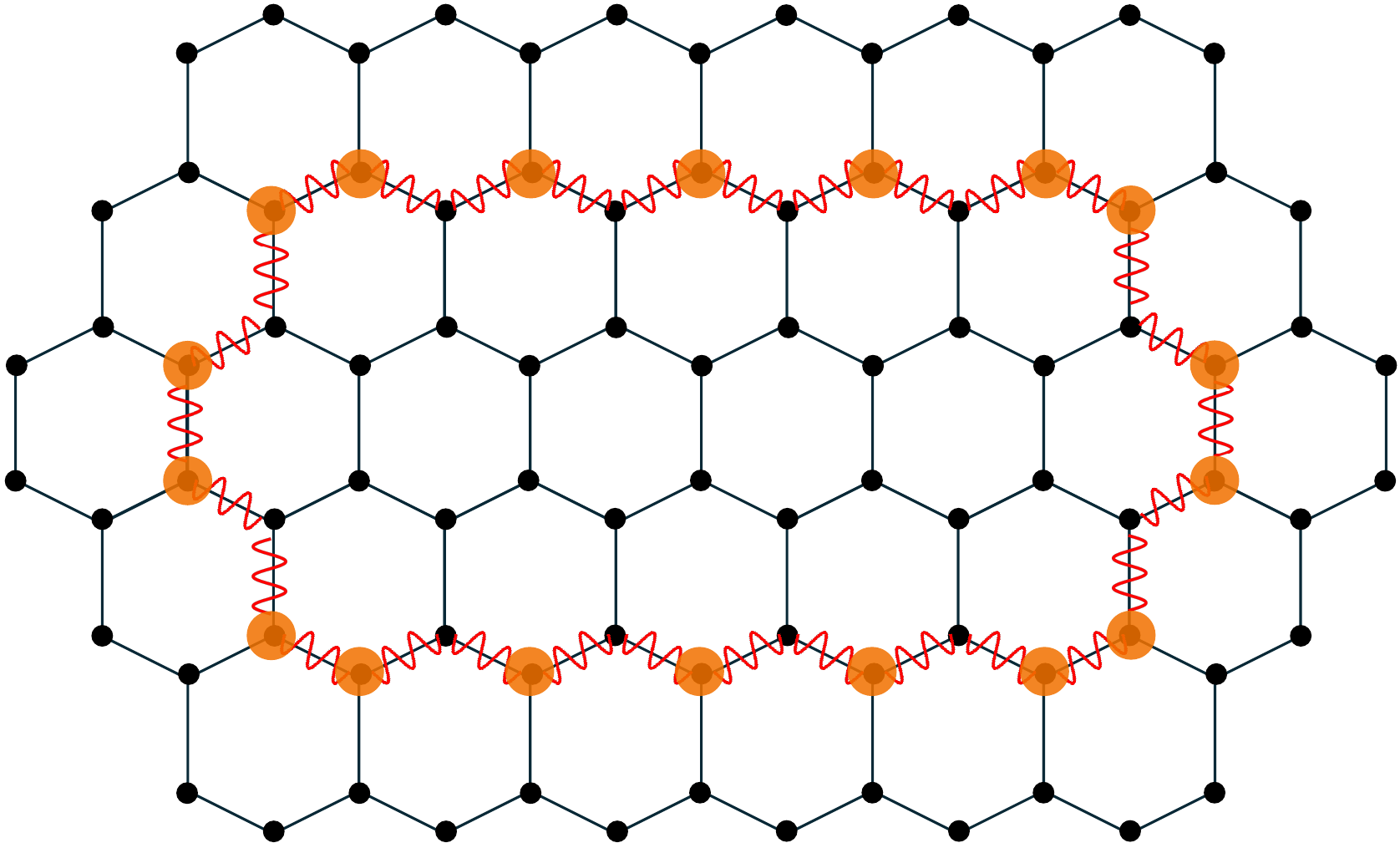}
\end{equation*}
where the red wiggling curve depicts the two-body CZ gates and the orange dots label the location of the single-site Pauli-Zs. It is also straightforward to see that $\ket{\phi}$ respects the global $\mathbb{Z}_2$ (0-form) symmetry generated by $\prod_{i \in B \cup C } X_i$.

These two symmetries carry a mixed anomaly, which can be shown by the algebra between their truncated symmetry actions. Specifically, we denote a truncated 0-form symmetry operator as $U_{\text{trunc}}$ and a truncated 1-form symmetry operator as $W_{\text{trunc}}$. They create a loop-like defect (green circle) and two point-like defects (red dots) in the lattice below: 

\begin{equation*}
\includegraphics[width=10cm]{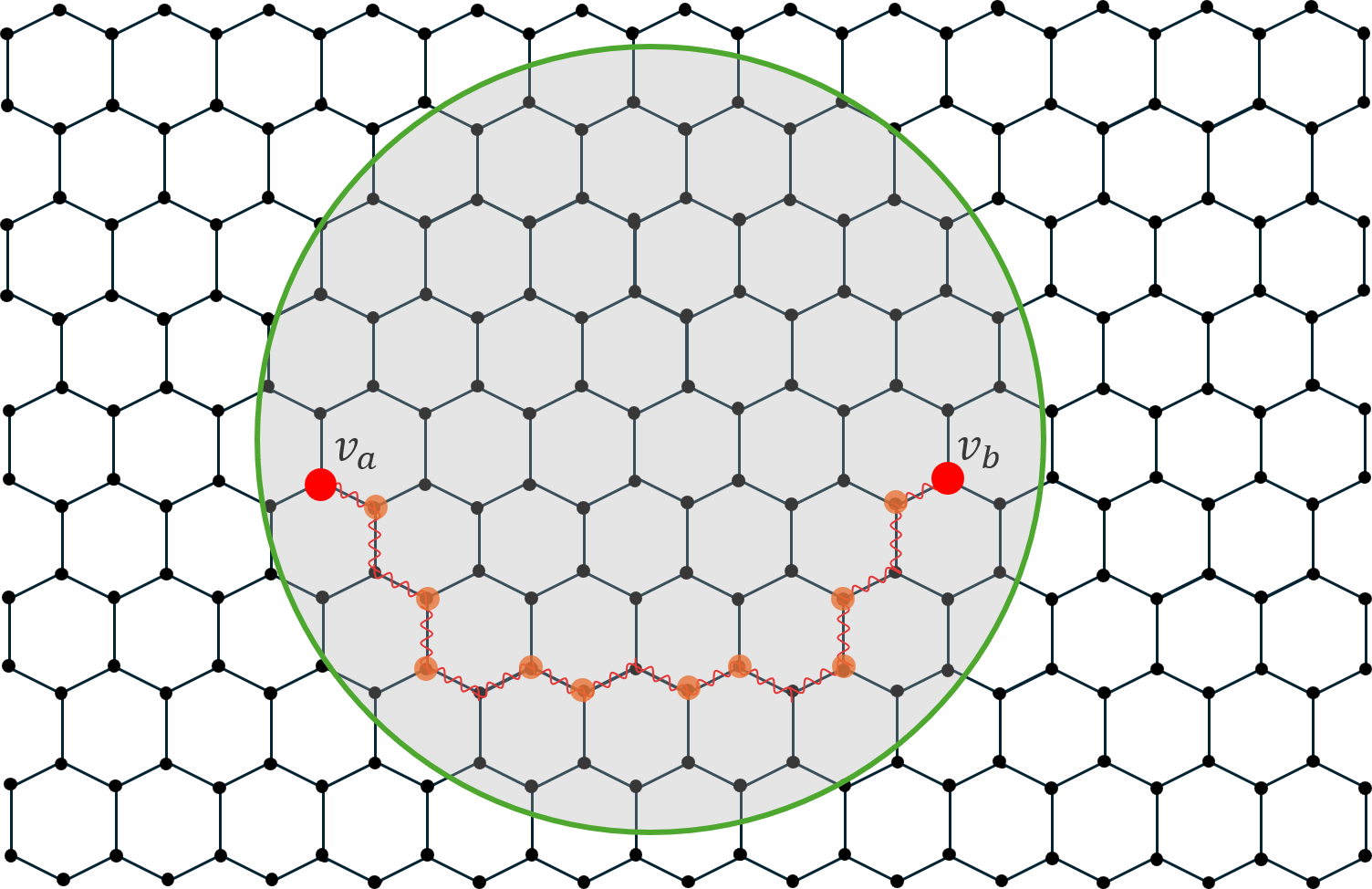}
\end{equation*}

We assume the two end points of the truncated string operator, $v_a$ and $v_b$ are far from each other, and both of them are far from the green circle. The braiding between the loop-like defect and the two point-like defects gives rise to
\begin{equation}
    \begin{aligned}
        U_{\text{trunc}}W_{\text{trunc}}U_{\text{trunc}}^{\dagger}W_{\text{trunc}}^{\dagger}&=\Big( \prod_{v\in \mathcal{R}}X_v\Big)\Big(\prod_{i=1} (-1)^{\frac{1}{6}}Z_{v_i} CZ_{v_i,v_{i+1}}\Big)\Big( \prod_{v\in \mathcal{R}}X_v\Big)^{\dagger}\Big(\prod_{i=1} (-1)^{\frac{1}{6}}Z_{v_i} CZ_{v_i,v_{i+1}}\Big)^{\dagger}\\
        &=Z_{v_a}Z_{v_b},
    \end{aligned}
    \label{eq:defect_braiding}
\end{equation}
up to a minus sign. Thus the operator $Z_{v_a}$ near endpoint $v_a$ can be seen as the braiding between a loop-like defect and one point-like defect. Furthermore, this operator and a loop-like defect have a nontrivial braiding phase,
\begin{equation}
    U_{\text{trunc}} Z_{v_a} U_{\text{trunc}}^{\dagger} Z_{v_a} = -1.
    \label{eq:anomaly_braiding_phase}
\end{equation}
There is a phase ambiguity in defining the local braiding operator near $v_a$, which will be discussed later. This nontrivial topological data between the symmetry defects indicates the mixed anomaly between the two symmetries \cite{wen2019emergent}. This mixed anomaly is analogous to the mixed anomaly between two 0-form $\mathbb{Z}_2$ symmetries in 1d discussed in Ref.\cite{else2014classifying,kawagoe2021anomalies,cheng2023lieb}, and the above braiding phase is generalized from the 3-cocycle $\omega(g,h,k)$ defined in Ref.\cite{else2014classifying}. We therefore expect $\ket{\phi}$ to be long-range entangled. A simple argument can be made as follows. 

Suppose the state $\ket{\phi}$ is a short-ranged entangled symmetric state, then $W_{\text{trunc}}\ket{\phi}=\Sigma^{\dagger}_{a}\Sigma'_{b}\ket{\phi}$, where $\Sigma^{\dagger}_{a}$ is a local unitary operator near $v_a$, and $\Sigma'_{b}$ is a local unitary near $v_b$. We redefine the truncation of the 1-form symmetry as
\begin{equation}
    \Tilde{W}_{\text{trunc}}=\Sigma_{a}W_{\text{trunc}},
\end{equation} 
such that $\Tilde{W}_{\text{trunc}}\ket{\phi}=\Sigma'_{b}\ket{\phi}$. With the modified truncation, the braiding between the two point-like defects and a loop-like defect is given by
\begin{equation}
    \begin{aligned}
        U_{\text{trunc}}\Tilde{W}_{\text{trunc}}U_{\text{trunc}}^{\dagger}\Tilde{W}_{\text{trunc}}^{\dagger}=U_{\text{trunc}}\Sigma_{a}W_{\text{trunc}}U_{\text{trunc}}^{\dagger}W_{\text{trunc}}^{\dagger}\Sigma^{\dagger}_{a}=B_a B_b,
    \end{aligned}
\end{equation}
where $B_a$ is a local unitary near $v_a$, and $B_b$ is a local unitary near $v_b$. Now we show the braiding between $B_a$ and the loop-like defect still gives a nontrivial phase,
\begin{equation}
\begin{aligned}
    &U_{\text{trunc}} B_a U_{\text{trunc}}^{\dagger} B_a \\
    &= U_{\text{trunc}}\left(U_{\text{trunc}}\Sigma_{a}W_{\text{trunc}}U_{\text{trunc}}^{\dagger}W_{\text{trunc}}^{\dagger}\Sigma^{\dagger}_{a}\right)_a U_{\text{trunc}}^{\dagger} \left(U_{\text{trunc}}\Sigma_{a}W_{\text{trunc}}U_{\text{trunc}}^{\dagger}W_{\text{trunc}}^{\dagger}\Sigma^{\dagger}_{a}\right)_a\\
    &= U_{\text{trunc}}\left(U_{\text{trunc}}\Sigma_{a}U_{\text{trunc}}^{\dagger}U_{\text{trunc}}W_{\text{trunc}}U_{\text{trunc}}^{\dagger}W_{\text{trunc}}^{\dagger}\right)_a \Sigma^{\dagger}_{a}U_{\text{trunc}}^{\dagger} \left(U_{\text{trunc}}\Sigma_{a}U_{\text{trunc}}^{\dagger}U_{\text{trunc}}W_{\text{trunc}}U_{\text{trunc}}^{\dagger}W_{\text{trunc}}^{\dagger}\right)_a \Sigma^{\dagger}_{a}\\
    &= U_{\text{trunc}}U_{\text{trunc}}\Sigma_{a}U_{\text{trunc}}^{\dagger}\left(U_{\text{trunc}}W_{\text{trunc}}U_{\text{trunc}}^{\dagger}W_{\text{trunc}}^{\dagger}\right)_a \Sigma^{\dagger}_{a}U_{\text{trunc}}^{\dagger} U_{\text{trunc}}\Sigma_{a}U_{\text{trunc}}^{\dagger}\left(U_{\text{trunc}}W_{\text{trunc}}U_{\text{trunc}}^{\dagger}W_{\text{trunc}}^{\dagger}\right)_a \Sigma^{\dagger}_{a}\\
    &=\Sigma_{a}U_{\text{trunc}}\left(U_{\text{trunc}}W_{\text{trunc}}U_{\text{trunc}}^{\dagger}W_{\text{trunc}}^{\dagger}\right)_a U_{\text{trunc}}^{\dagger} \left(U_{\text{trunc}}W_{\text{trunc}}U_{\text{trunc}}^{\dagger}W_{\text{trunc}}^{\dagger}\right)_a \Sigma^{\dagger}_{a}\\
    &=-\Sigma_{a}\Sigma^{\dagger}_{a}=-1.
\end{aligned}
\label{eq:anomalous_braiding}
\end{equation}
In the above equation, we use $(\cdots)_a$ to denote the part of the operator $(\cdots)$ that is near $v_a$. From the second line to the third line, we took $\Sigma_a^{\dagger}$ out of the bracket $(\cdots)_a$ because it is already a local unitary near $v_a$. From the third line to the fourth line, we took $U_{\text{trunc}}\Sigma_a U_{\text{trunc}}^{\dagger}$ outside of the bracket, since it is a local unitary near $v_a$. From the fourth line to the fifth line, we used $U_{\text{trunc}}^2=1$. From the fifth line to the sixth line, we used Eq.\ref{eq:defect_braiding} and Eq.\ref{eq:anomaly_braiding_phase}.

On the other hand, we recall that $\Tilde{W}_{\text{trunc}}\ket{\phi}=\Sigma'_{b}\ket{\phi}$, and $U_{\text{trunc}}\ket{\phi}=\mathcal{L}_{\partial \mathcal{R}}\ket{\phi}$, where $\mathcal{L}_{\partial \mathcal{R}}$ is a unitary near the boundary of region $\mathcal{R}$. When $v_a$ and $v_b$ are far from each other and are both far away from $\partial \mathcal{R}$, the two operators commute up to some local unitaries near $v_a$ and $v_b$,
\begin{equation}
    U_{\text{trunc}}\Tilde{W}_{\text{trunc}}U_{\text{trunc}}^{\dagger}\Tilde{W}_{\text{trunc}}^{\dagger}=B_a B_b.
\end{equation}
Importantly, since both $U_{\text{trunc}}$ and $\Tilde{W}_{\text{trunc}}$ locally stabilize state $\ket{\phi}$, the local unitary $B_a$ also stabilizes $\ket{\phi}$. Thus we have
\begin{equation}
    \begin{aligned}
        &U_{\text{trunc}}B_a U_{\text{trunc}}^{\dagger}B_a\ket{\phi}\\
        &=U_{\text{trunc}}B_a U_{\text{trunc}}^{\dagger}\ket{\phi}\\
        &=U_{\text{trunc}}B_a \mathcal{L}_{\partial\mathcal{R}}\ket{\phi}\\
        &=U_{\text{trunc}} \mathcal{L}_{\partial\mathcal{R}}B_a\ket{\phi}\\
        &=U_{\text{trunc}} \mathcal{L}_{\partial\mathcal{R}}\ket{\phi}\\
        &=U_{\text{trunc}} U_{\text{trunc}}^{\dagger}\ket{\phi}\\
        &=\ket{\phi}.
    \end{aligned}
\end{equation}
From the first line to the second line, we used the fact that $B_a$ stabilizes the state. From the third line to the fourth line, we used that $B_a$ and $\mathcal{L}_{\partial\mathcal{R}}$ commute, since they do not overlap. This equation directly contradicts the result in Eq.\ref{eq:anomalous_braiding}.

We note that the definition of $B_a$ above is ambiguous up to a phase, i.e., we can always assign an extra phase $i$ to $B_a$, and phase $-i$ to $B_b$, such that the braiding $U_{\text{trunc}} B_a U_{\text{trunc}}^{\dagger} B_a$ becomes trivial. Nonetheless, we can define a different braiding operator as $C_a=(U_{\text{trunc}}W_{\text{trunc}}^{\dagger}U_{\text{trunc}}^{\dagger}W_{\text{trunc}}^{\dagger})_a$, and the following phases
\begin{equation}
    \begin{aligned}
        C_a W_{\text{trunc}}B_a & W_{\text{trunc}}\\
        &=\left(U_{\text{trunc}}W_{\text{trunc}}^{\dagger}U_{\text{trunc}}^{\dagger}W_{\text{trunc}}^{\dagger}\right)_a W_{\text{trunc}}\left(U_{\text{trunc}}W_{\text{trunc}}U_{\text{trunc}}^{\dagger}W_{\text{trunc}}^{\dagger}\right)_a W_{\text{trunc}} ,\\
        U_{\text{trunc}}W_{\text{trunc}}^{\dagger} & W_{\text{trunc}}^{\dagger} U_{\text{trunc}}^{\dagger}B_a C_a^{\dagger}\\
        &= U_{\text{trunc}}W_{\text{trunc}}^{\dagger}W_{\text{trunc}}^{\dagger} U_{\text{trunc}}^{\dagger}\left(U_{\text{trunc}}W_{\text{trunc}}U_{\text{trunc}}^{\dagger}W_{\text{trunc}}^{\dagger}\right)_a \left(W_{\text{trunc}}U_{\text{trunc}}W_{\text{trunc}}U_{\text{trunc}}^{\dagger}\right)_a.
    \end{aligned}
\end{equation}
For our original definition of truncation, we can let $B_a=C_a=Z_{v_a}$, then the above two phases are trivial. Following a similar derivation as in Eq.\ref{eq:anomalous_braiding}, we can show that the above two phases are invariant under the redefinition of truncation $\Tilde{W}_{\text{trunc}}$. No matter how we redefine the local braiding operators $B_a$ and $C_a$ by a phase, at least one of the above quantities would be nontrivial. On the other hand, we can show that after defining the truncated operator as $\Tilde{W}_{\text{trunc}}$, all of them stabilize the state $\ket{\phi}$. As a result, the symmetric state $\ket{\phi}$ should be long-range entangled.

\subsection{Wavefunction Overlap}\label{appendix:overlap} Here we calculate the overlap between $\ket{\psi}$ and $W_{\Gamma_x}\ket{\psi}$. By mapping to the Ashkin-Teller model on a triangular lattice, we argue the overlap decays exponentially small in the system size. To start, 
 
\begin{equation}
\frac{\bra{\psi} W_{\Gamma_x} \ket{\psi}}{\bra{\psi}\ket{\psi}  }=  \frac{\bra{+} W_{\Gamma_x} \prod_p(1+U_p) \ket{+}   }{\bra{+} \prod_p(1+U_p)  \ket{+} } =\frac{\sum_{\mathcal{C'}}  \bra{+} \ket{\text{SPT}}_{\mathcal{C'}}}{ \sum_{\mathcal{C}}  \bra{+} \ket{\text{SPT}}_{\mathcal{C}}},
\end{equation}
where $\mathcal{C}$ is a contractible loop, and $\mathcal{C'}$ is non-contractible along $\hat{x}$ direction. The overlap between a $L$-qubit $\ket{+}$ state and a $L$-qubit $\mathbb{Z}_2\times\mathbb{Z}_2$ SPT cluster state (with $L$ being even) is $2^{-\frac{L}{2}+1}$, and therefore  $\bra{+} \ket{\text{SPT}}_{\mathcal{C}} =2^{-\frac{\abs{\mathcal{C}}}{2}  }2^{N_\mathcal{C}}$, where $N_{\mathcal{C}}$ is the number of disconnected loops in $\mathcal{C}$. 

To compute $\bra{+}\ket{\text{SPT}}_{\mathcal{C'}}$, we consider the system sizes in the two spatial 
directions to be even by even, in which case, 
the non-contractible loop $\mathcal{C}'$ must have even number of qubits and $\bra{+} \ket{\text{SPT}}_{\mathcal{C'}} = 2^{-\frac{\abs{\mathcal{C'}}}{2}} 2^{N_\mathcal{C'}}$. As a result, one finds

\begin{equation}\label{eq:overlap}
\frac{\bra{\psi} W_{\Gamma_x} \ket{\psi}}{\bra{\psi}\ket{\psi}  }=\frac{\sum_{\mathcal{C'}}  \left(\frac{1}{\sqrt{2}} \right)^{\abs{\mathcal{C'} } } 2^{ N_{\mathcal{C}'}}}{ \sum_{\mathcal{C}} \left(\frac{1}{\sqrt{2}}\right)^{\abs{\mathcal{C} } }2^{ N_{\mathcal{C}}}   }. 
\end{equation}

This is the ratio of the two 
partition functions of the $O(2)$ loop models, one with only contractible loops and one with only non-contractible loops.

It is known that the $O(n)$ loop model on the honeycomb lattice can be exactly mapped to the $n$-color Ashkin-Teller (AT) model at the infinite-coupling limit on a triangular lattice \cite{deng2006loop_mapping,HUANG2013_loop}. Here we review the mapping for $n=2$, and based on which, show that the overlap in Eq.\ref{eq:overlap} relates to the free energy cost of inserting a line defect in the two-color AT model at a critical point, therefore indicating the exponential decay of the wavefunction overlap.

Consider a triangular lattice where each lattice site has two flavors of Ising spins, the partition function of the corresponding AT model reads 

\begin{equation}
Z_{\text{AT}} = \sum_{\sigma,\tau} e^{-H_{\text{AT}}}  \qquad \text{with } H_{\text{AT}} = - J_2 \sum_{\expval{ij}} (\sigma_i\sigma_j +\tau_i\tau_j  )  - J_4 \sum_{ \expval{ij} } \sigma_i \sigma_j \tau_i\tau_j
\end{equation}
Below we show that at the infinite-coupling limit: $J_2 \to \infty, J_4 \to -\infty $ with $J_2 +J_4 =J$ fixed at a finite value, the AT model can be mapped to the $O(2)$ model on the honeycomb lattice using a low-temperature expansion. On a given bond $\expval{ij}$ on the triangular lattice, there are three cases:\\

\noindent (i) $\sigma_i \sigma_j=\tau_i \tau_j=1$: the corresponding Boltzmann weight is $e^{2J_2 +J_4}$. \\

\noindent (ii) only one disagreement on the bond $\expval{ij}$, i.e. $\sigma_i \sigma_j=1= - \tau_i \tau_j$ or $-\sigma_i \sigma_j=1=\tau_i \tau_j$, the corresponding Boltzmann weight is 
$e^{-J_4}$. \\

\noindent (iii) both flavors of the spins disagree, i.e. $\sigma_i \sigma_j=\tau_i \tau_j=-1$, the corresponding Boltzmann weight is $e^{ -2J_2 + J_4}$. \\

If one normalizes the Boltzmann weight w.r.t the weight in the case (i), one has the normalized weights:

Case (i): 1, Case (ii): $e^{-2J_2 - 2J_4} \equiv  e^{-2J}$, and  Case (iii): $e^{-4J_2}$.

Therefore, by taking the limit $J_2 \to \infty, J_4 \to -\infty $ with $J_2 +J_4 =J$, one only needs to consider the case (i) and (ii). The probability of having case (iii) would be zero, which implies that the domain walls of the $\sigma$ spins and $\tau$ spins cannot overlap, e.g.  see below for allowed configurations: 
\begin{equation*}
\includegraphics[width=7cm]{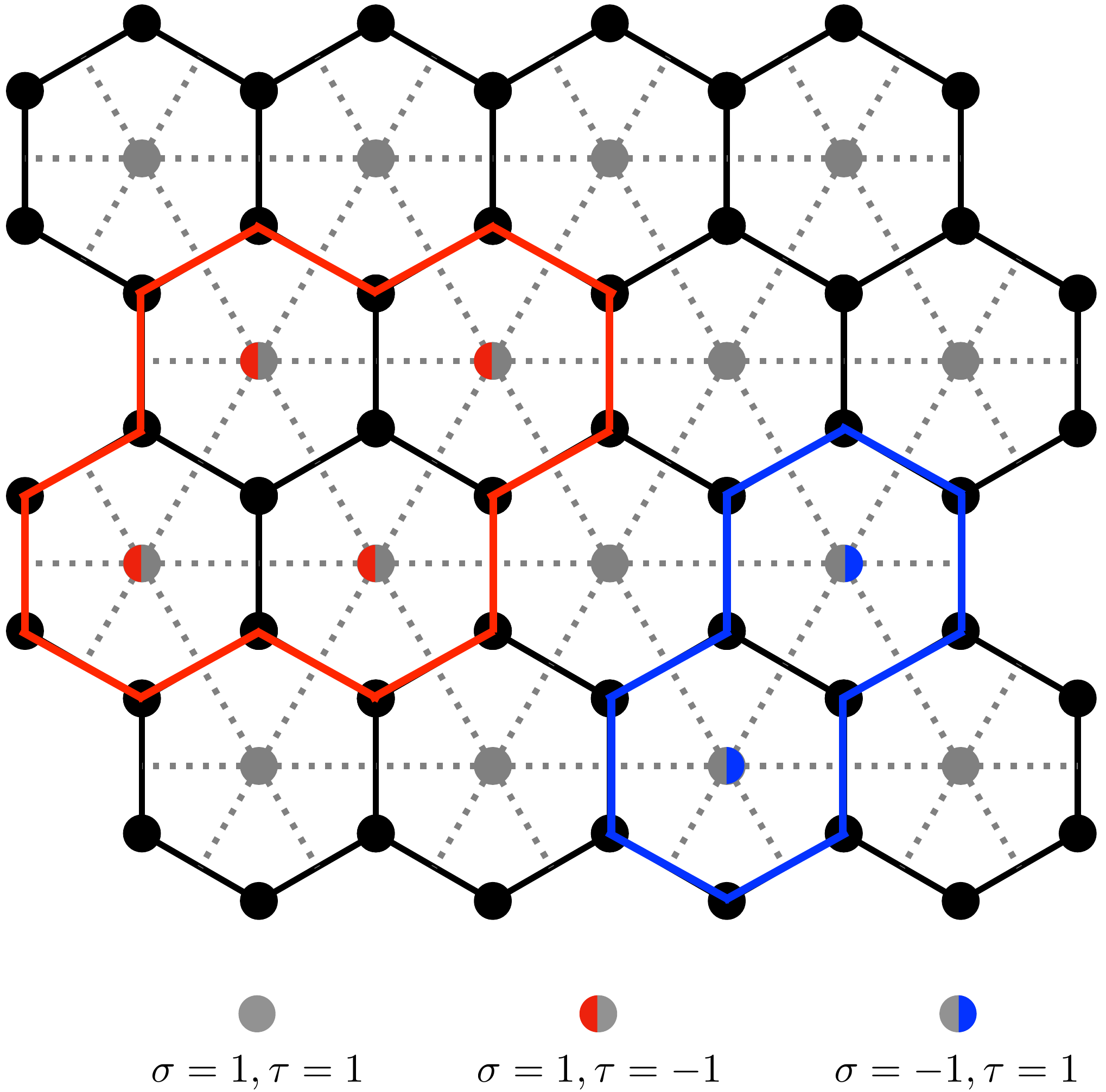}
\end{equation*}

Correspondingly, when considering the low-temperature expansion by flipping spins, the partition function (up to a constant) of the infinite-coupling limit Ashkin-Teller (ICLAT) model is

\begin{equation}\label{append:mapping_loop}
Z_{\text{ICLAT}} =   \sum_{\mathcal{C}} x^{\abs{ \mathcal{C}}} 2^{N_{\mathcal{C}}}, 
\end{equation}
where $\mathcal{C}$ is any contractible loop, with the loop tension $x\equiv e^{-2J}$.  The factor $2^{N_{\mathcal{C}}}$ appears since each loop can be the domain wall of either one of the two flavors of spins. Eq.
\ref{append:mapping_loop} is exactly the partition function of the $O(2)$ loop model.

The phase diagram of the infinite-coupling AT model is studied in Ref.\cite{HUANG2013_loop}; for $x=e^{-2J}<x_c= \frac{1}{\sqrt{2}}$, both $\sigma, \tau$ spins are in the ferromagnetic ordered phase, while for $x>x_c$, the spins are in the critical phase. $x_c$ is the critical point described by a Berezinskii-Kosterlitz-Thouless (BKT) transition.

Alternatively, the model can be written as two flavors of the Ising models subject to the constraint:
\begin{equation}
Z_{\text{ICLAT}} = \sum_{\sigma, \tau } e^{J\sum_{\expval{ij}}  (\sigma_i \sigma_j + \tau_i \tau_j)  } \prod_{\expval{ij}} \delta_{ij} (\sigma_i \sigma_j \neq -1~\&~ \tau_i \tau_j \neq -1),
\end{equation}
where the delta function enforces the constraint that on every bond, the domain wall of $\sigma$ and $\tau$ cannot exist simultaneously. For large $J$, one expects both $\sigma$ and $\tau$ to be in the ordered phase. When decreasing $J$ (increasing temperature), since the domain wall of $\sigma$ and $\tau$ cannot be present at the same time, the spins cannot be completely disordered, i.e. the domain wall cannot condense at the same time. This gives an intuitive reasoning for the emergence of the critical phase.

Therefore, the wavefunction overlap relates to the free-energy cost of inserting a non-contractible line defect: 
\begin{equation} 
\frac{\bra{\psi} W_{\Gamma_x} \ket{\psi}}{\bra{\psi}\ket{\psi}  }= \frac{Z'_{\text{ICLAT}}}{Z_{\text{ICLAT}}}, 
\end{equation} 
where $Z'_{\text{ICLAT}}$ is the AT model at the critical point with a line defect obtained by flipping the sign of spin coupling along a non-contractible loop of size $L$. At the critical point, one expects  $\frac{Z'_{\text{ICLAT}}}{Z_{\text{ICLAT}}} \sim  e^{-\alpha L}$ at the leading order with $\alpha$ being an $O(1)$ constant so the wavefunction overlap vanishes in the thermodynamic limit \cite{oshikawa_1997_cft_defect}.

\subsection{Two-point functions}\label{appendix:2point}
Here we compute the two-point functions   $Z_{i_A}Z_{j_A}$ on the sublattice A: 
\begin{equation}
    \begin{split}
        \bra{\psi} Z_{i_A}Z_{j_A} \ket{\psi} &= \frac{\bra{+} \prod_p \frac{1 + U_p}{2} Z_{i_A}Z_{j_A} \prod_p \frac{1 + U_p}{2}\ket{+}}{\bra{+} \prod_p \frac{1 + U_p}{2} \prod_p \frac{1 + U_p}{2}\ket{+}} \\
         &= \frac{\bra{+} \prod_p \frac{1 + U_p}{2} Z_{i_A}Z_{j_A} \ket{+}}{\bra{+} \prod_p \frac{1 + U_p}{2} \ket{+}} \\
         &= \frac{\sum_{\mathcal{C}}\bra{+}_{\Bar{\mathcal{C}}}\bra{\mathrm{SPT}}_{\mathcal{C}} Z_{i_A}Z_{j_A} \ket{+}_{\mathcal{C}} \ket{+}_{\Bar{\mathcal{C}}}}{\sum_{\mathcal{C}}\bra{+}_{\Bar{\mathcal{C}}}\bra{\mathrm{SPT}}_{\mathcal{C}} \ket{+}_{\mathcal{C}} \ket{+}_{\Bar{\mathcal{C}}}}
    \end{split}
\end{equation}
where from the second to the third line, we expand the product over plaquette $p$.
Now we can consider applying $Z_{i_A}Z_{j_A}$ to the right-hand-side $\ket{+}$ state. If $Z_{i_A}Z_{j_A}$ has support on $\Bar{\mathcal{C}}$, then the inner product $\bra{+}_{\Bar{\mathcal{C}}}Z_{i_A}Z_{j_A} \ket{+}_{\Bar{\mathcal{C}}}$ vanishes. The only case when it is non-vanishing is $Z_{i_A}Z_{j_A}$ acting on the same loop $\gamma$ in the loop configuration $\mathcal{C}$. If $Z_{i_A}Z_{j_A}$ acts on different loops $\gamma_1$ and $\gamma_2$, then the inner product $\bra{\mathrm{SPT}}_{\gamma_1} Z_{i_A} \ket{+}_{\gamma_1}$ also vanishes. The overlap between a $L$-qubit $\ket{+}$ state and a $L$-qubit $\mathbb{Z}_2\times\mathbb{Z}_2$ SPT cluster state (with $L$ being even) is $2^{-\frac{L}{2}+1}$. Therefore, the two-point function can be expressed as 
\begin{equation}
 \bra{\psi} Z_{i_A}Z_{j_A} \ket{\psi} = \frac{\sum_{\mathcal{C'}} (\frac{1}{\sqrt{2}})^{\abs{\mathcal{C'}}} 2^{N_{\mathcal{C}'}}}{\sum_{\mathcal{C}} (\frac{1}{\sqrt{2}})^{\abs{\mathcal{C}}} 2^{N_{\mathcal{C}}}},
\end{equation}
where $\mathcal{C'}$ is the loop configuration where there always exists a loop $\gamma$ connecting $i_A$ and $j_A$. This is exactly the 2-leg watermelon correlator in the $O(2)$ loop model with at the critical loop tension $K_c = \frac{1}{\sqrt{2}}$,  which is known to decay as \cite{duplantier1989two}.
\begin{equation}
     \bra{\psi} Z_{i_A}Z_{j_A} \ket{\psi} \sim \frac{1}{\abs{i_A - j_A}},
\end{equation}

\section{SymTFT description}\label{append:sym_tft}
The SymTFT/topological holography framework is known as an insightful way to describe quantum systems with symmetries~\cite{ji2020categorical,kong2020algebraic,gaiotto2021orbifold,lichtman2021bulk,freed2022topological,freed2022topologicalsym,moradi2023topological,bhardwaj2023categorical,bhardwaj2023gapped,huang2023topological,chatterjee2023symmetry}. The core idea of SymTFT is to encode the symmetry data of a quantum system into the topological order in a one-dimensional-higher bulk. This is illustrated with a sandwich picture, where the bulk topological order has a gapped boundary (reference/topological boundary) on the left and another (dynamical/physical) boundary on the right. The reference boundary realizes the specific symmetry data given from the bulk topological order, whereas the dynamical boundary is given from the details of the theory, and is potentially not topological. 

In the 1d model we discussed, the $\mathbb{Z}_2^3$ symmetry given by $U_e= \prod_i X_{2i}$, $U_o = \prod_i X_{2i+1}$, and the non-onsite $U_{\text{CZ}}=\prod_i CZ_{i,i+1}$  together form a type-III mixed-anomaly. The 2d bulk topological order for this anomalous symmetry is a twisted quantum double, which is also equivalent to a $D_8$ (the dihedral group of order 8) quantum double~\cite{propitius1995topological}. The reference boundary on the left corresponds to condensing all the abelian anyons (charges) in the topological order, given by the Lagrangian algebra $1+e_1+e_2+e_3+e_1e_2+e_1e_3+e_2e_3+e_1e_2e_3$~\cite{lan2015gapped}. The topological symmetry defects in this quasi-1d system are the insertions of $m_1$, $m_2$ and $m_3$ lines in the bulk topological order. The spontaneously breaking phase of one of the $\mathbb{Z}_2$ symmetries is a gapped phase, which corresponds to the choice of a gapped dynamical boundary on the right, given by Lagrangian algebra $1+e_1+m_2+m_3+m_{23}$~\cite{lan2015gapped}.

\begin{figure}[h!]
    \centering
    \includegraphics[width=0.45\linewidth]{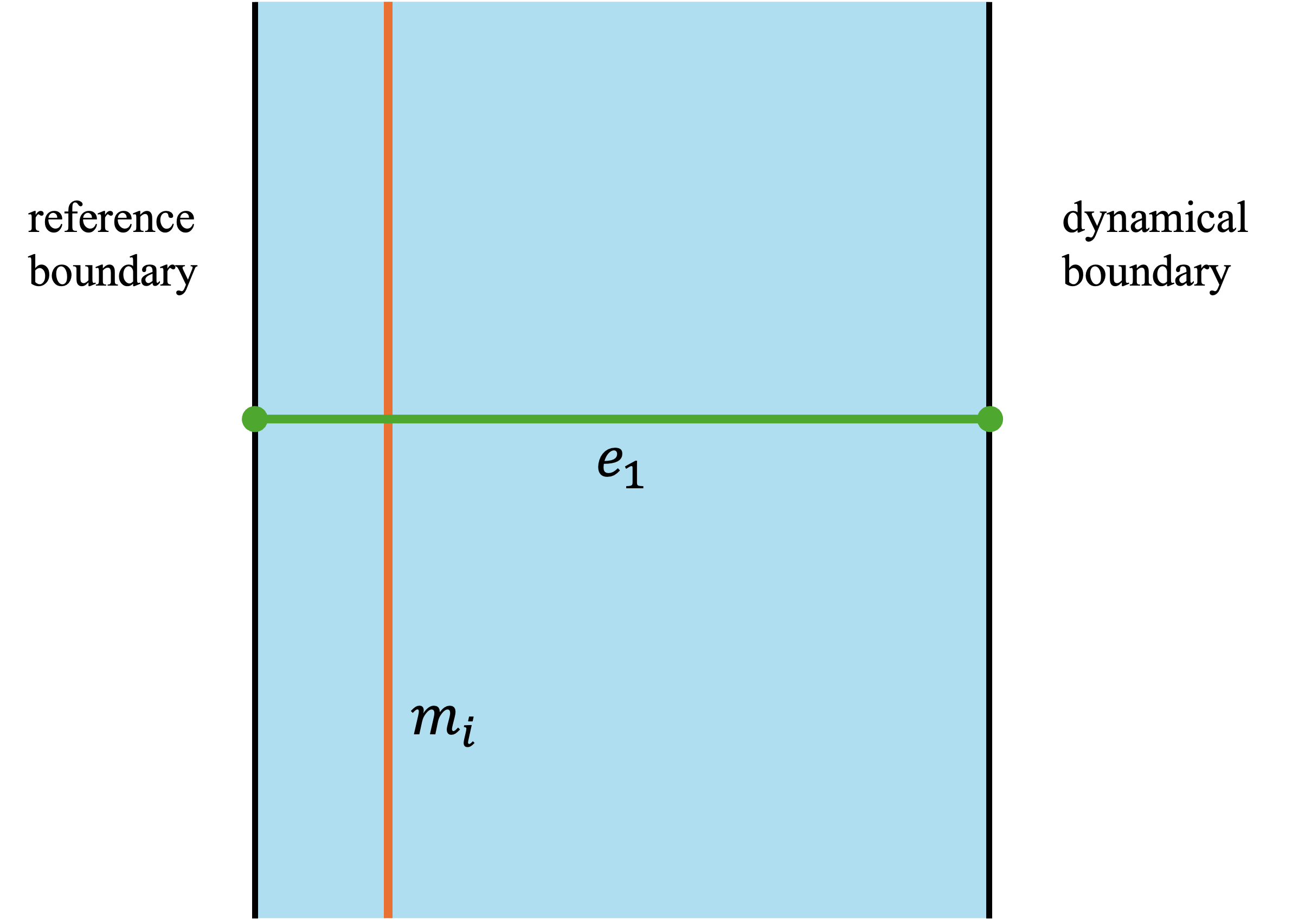}
    \caption{The bulk topological order is a $\mathbb{Z}_2^3$ type-III twisted quantum double. The left boundary is the reference/topological boundary, given by Lagrangian algebra $1+e_1+e_2+e_3+e_1e_2+e_1e_3+e_2e_3+e_1e_2e_3$. The symmetry of the quasi-1d system is given from this reference boundary, i.e. by inserting the $m_i$ lines denoted in orange. The right boundary is the dynamical/physical boundary, given by Lagrangian algebra $1+e_1+m_2+m_3+m_{23}$. The local operator denoted in green is an $e_1$ line that ends on both boundaries. It is charged under $m_1$, thus is an order parameter indicating the spontaneous breaking of this $\mathbb{Z}_2$ symmetry, or the emergent $\mathbb{Z}_2$ 1-form in this symmetry-breaking phase.} 
    \label{fig:symTFT}
\end{figure}

In choosing the two boundaries in the sandwich picture, there is an emergent 1-form symmetry in the system, which is given by the $e_1$ lines ending on both boundaries, denoted as a horizontal green line in the above figure. It braids non-trivially with $m_1$ lines, and braids trivially with $m_2$, $m_3$ lines. Therefore, this operator can be understood as the order parameter for the $\mathbb{Z}_2$ symmetry breaking, and the emergent $\mathbb{Z}_2$ 1-form operator in this symmetry breaking phase~\cite{ji2020categorical,kong2020algebraic,huang2023topological}.

In the symmetry-breaking phase of the non-onsite $U_{CZ}$, we have a co-existence of trivial/SPT phases under two onsite $\mathbb{Z}_2$ symmetries, which is mixed-anomalous of type-III. As we have shown in the main text, there is no unitary operator $V$ charged under $U_{CZ}$, i.e. $U_{CZ}VU_{CZ}^{\dagger}=-V$. While this seems to suggest the absence of an emergent 1-form symmetry in this non-onsite symmetry-breaking phase, hence creating a tension in the SymTFT description, there is still a possibility that such a 1-form symmetry may only emerge in IR \cite{pace_2024}, thereby being consistent with the framework of SymTFT. A deeper understanding of this potential emergent 1-form symmetry is an interesting question for the future.  

In the following, we give two closely related models with the same mixed-anomalous $\mathbb{Z}_2^3$ symmetry but in the SSB phase of one of the onsite $\mathbb{Z}_2$ symmetries, which manifests the above SymTFT picture.

The first model is given by 
\begin{equation}
    H=-\sum_i Z_{2i-1}X_{2i}Z_{2i+1}-\sum_i X_{2i},
\end{equation}
which has the same $\mathbb{Z}_2^3$ symmetry, but in the broken phase of the onsite $\mathbb{Z}_2$ symmetry $U_o$. The two ground states $\ket{\psi_0}=\ket{0\cdots0}_{odd}\otimes \ket{+\cdots+}_{even}$ and $\ket{\psi_1}=\ket{1\cdots 1}_{odd}\otimes \ket{+\cdots +}_{even}$
are both product states. We can also regard it as a trivial/SPT co-existing phase of $U_o$ and $U_{CZ}$ symmetry since $\ket{\psi_0}$ ($\ket{\psi_1}$) can be regarded as an SPT state obtained from decorating trivial (nontrivial) $U_{CZ}$ charge on the $U_e$ domain walls. Specifically, both states contain $\ket{+\cdots +}_{even}\propto\sum_{s_{2i}=0,1}\ket{s_2,s_4,\cdots}$ on the even sublattice. This can be understood as a condensation of $U_e$ domain walls, where a $U_e$ domain wall between site $2i$ and site $2i+2$ refers to the configurations $s_{2i}\neq s_{2i+2}$. Since $\ket{\psi_1}$ always has $s_{2i+1}=1$ on the site $2i+1$, applying $\prod_i CZ_{i,i+1}$ gives rise to a $-1$ phase on this domain wall (i.e. a non-trivial $U_{CZ}$-charge decoration). In contrast, $\ket{\psi_1}$ has $s_{2i+1}=0$ on the site $2i+1$, so applying $\prod_i CZ_{i,i+1}$ gives the $+1$ sign (a trivial charge) on the $U_e$ domain walls. We note that for this model, there is an emergent 1-form symmetry in the ground subspace, the operator of which is exactly the order parameter $Z_{2i+1}Z_{2j+1}$ for the onsite $U_o$ symmetry breaking.

The second model is obtained by introducing an ancilla qubit at state $\ket{+}_{\tau}$ on each even site of the spin chain, which is similar to the setting in Append.\ref{append:ancilla}. We consider the diagonal symmetry $\mathbb{Z}_2^{diag}$ between the non-onsite $U_{CZ}$ and the onsite $\prod_i \tau^x_{2i}$ symmetry on the ancillae. The symmetry operator is given by
\begin{equation}
    U_{diag}=\prod_i CZ_{i,i+1}\tau^x_{2i},
\end{equation}
which is non-onsite. As shown in Append.\ref{append:ancilla}, the non-onsite operator $U_{diag}$ becomes onsite after the conjugation of a unitary $V$ in Eq.\ref{eq:conjugation_V}. In fact, the spontaneously breaking phase of this onsite $\prod_i \tau^x_{2i}$ symmetry can be obtained from a trivially symmetric phase with onsite $D_4$ symmetry, via a Kramers-Wannier transformation for the $\mathbb{Z}_2$ center. The fixed-point ground states are simply given by
\begin{equation}
    \ket{\psi'_{\pm}}=\ket{+}^{\otimes N}\otimes \frac{1}{\sqrt{2}}\left(\ket{0...0}_{\tau}\pm\ket{1...1}_{\tau}\right).
\end{equation}
Therefore, the ground states of the non-onsite $\mathbb{Z}_2^{diag}$ symmetry breaking phase are
\begin{equation}
    \ket{\psi_{\pm}}=V^{\dagger}\ket{\psi'_{\pm}}=\frac{1}{\sqrt{2}}\left(\ket{+}\otimes \ket{0...0}_{\tau}\pm\ket{c}\otimes\ket{1...1}_{\tau}\right).
\end{equation}
There is an emergent 1-form symmetry in this phase, and the operator is $\tau_{2i}^z \tau_{2j}^z$.\\

 \twocolumngrid 

\bibliography{ref}

\end{document}